\documentclass[prb,twocolumn,amsmath,amssymb,showpacs]{revtex4}
\usepackage{graphicx}
 
\begin{document}

%
%

\title{Multi-Orbital Lattice Model for (Ga,Mn)As and Other Lightly 
Magnetically Doped Zinc-Blende-Type Semiconductors}

\author{Adriana Moreo$^{1}$, Yucel Yildirim$^{1,2}$, and Gonzalo Alvarez$^3$}
 
\address{$^1$Department of Physics and Astronomy,University of Tennessee,
Knoxville, TN 37966-1200 and
\\Materials Science and Technology Division
\\Oak Ridge National Laboratory,Oak Ridge,
TN 37831-6032,USA, }
 
\address{$^2$Department of Physics and  National High Magnetic Field Lab,
\\ Florida State University, Tallahassee, FL 32306, USA,}

\affiliation{$^3$ Computer Science and Mathematics Division and
Center for Nanophase Materials Sciences, Oak Ridge
National Laboratory, Oak Ridge, TN 37831-6032, USA.}

\date{\today}

\begin{abstract}

We present a Hamiltonian in real space which is well suited
to study numerically the behavior of holes introduced in III-V semiconductors
by Mn doping when the III$^{3+}$ ion is replaced by Mn$^{2+}$.
We consider the actual lattice with the diamond structure. Since the focus 
is on light doping by acceptors, a bonding combination of III and V p-orbitals 
is considered since the top of the valence band, located at the $\Gamma$ 
point, has p character in these materials. As a result, an effective model in 
which the holes hop between the sites of an fcc lattice is obtained. 
We show that around the $\Gamma$ point in momentum space the Hamiltonian for 
the undoped case is identical to the one for the Luttinger-Kohn model. 
The spin-orbit interaction is included as well as the on-site interaction
between the spin of the magnetic impurity and the spin of the doped holes. 
The effect of Coulomb interactions between Mn$^{2+}$ and holes, as well as 
Mn$^{3+}$ doping are discussed.
Through large-scale Monte Carlo simulations on a Cray XT3 supercomputer, 
we show that this model reproduces many experimental results for 
${\rm Ga_{\it 1-x}Mn_{\it x}As}$ and ${\rm Ga_{\it 1-x}Mn_{\it x}Sb}$, 
and that the Curie 
temperature does not increase monotonically with $x$. The cases of Mn doped 
GaP and GaN, in which Mn$^{3+}$ is believed to play a role, are also studied. 

\end{abstract}
 
 
\maketitle

\section{Introduction}

The discovery of high Curie temperatures ($T_C$) in some diluted magnetic 
semiconductors (DMS)\cite{OHN96,dietl00,Pot} has created a renewed interest 
in  
the study of the properties of these materials due to their potential role in 
spintronics devices if above room temperature $T_C$'s are achieved.\cite{Zutic}

When a III-V compound is magnetically doped the added holes can gain kinetic 
energy by moving in the bands formed by the hybridized orbitals or magnetic 
energy by interacting with the spin of the localized magnetic impurities. The 
kinetic term induces delocalization and it is usually studied in momentum 
space; the magnetic term tends to localize the carriers and it is natural to 
study it in real space. In the limit of large Mn doping the kinetic term 
should 
prevail while in the very diluted regime the magnetic term has to be 
accurately considered. Most studies of DMS are based in one of these two 
extreme
limits. However, in order to understand the intermediate region 
between both extremes powerful non-perturbative techniques are needed. 
The goal 
of this work is to provide such a tool. 

The band 
structure of Zinc-Blende type semiconductors has been accurately obtained 
using a variety of different approaches such as pseudopotential methods, 
tight-binding techniques, and by careful consideration of the symmetries 
involved through the application of group theory.\cite{cardona} 
However, many 
of these calculations involve a large number of orbitals per site making them 
unsuitable for simulations of doped systems with present day computers. 
The successful 
Luttinger-Kohn model,\cite{KL} that describes the top of the 
valence band of semiconductors with diamond structure is very useful to study
hole doping in cases of strong hybridization. When the doped holes are 
not localized they go into the valence band and a formulation of the problem 
in momentum space is appropriate. This approach, known as the valence band 
or hole-fluid scenario, has already been applied to
DMS\cite{macdonald,Dietl} providing results that would be valid in a regime in
which the doped holes are uniformly distributed in the crystal and the 
randomness in the
impurity distribution becomes irrelevant. On the other hand, it is well known 
that in 
the limit of strong dilution the doped holes will have a tendency to become
localized at the Mn impurities. In this extreme regime holes move by hopping 
through overlapping orbitals and this behavior is better described in real 
space. The study of this limit has been carried out only at a phenomenological 
level using tight binding Hamiltonians that, while taking into account the 
random position of the impurities, do not provide a realistic band description
of the bands resulting from the overlaps of the atomic 
orbitals involved.\cite{Gonz,Mat,Ken,Sarma,Schl,Xu} 

The goal of this paper is to provide a realistic, yet simple, 
tight-binding-like Hamiltonian, in real space, that reproduces the top of the 
valence band of Zinc-Blende type semiconductors with the smallest possible 
number of degrees of freedom per site, and that handles the interactions 
between 
the doped holes and the randomly distributed magnetic impurities.
It will be shown that six orbitals per site are needed to capture experimental 
properties of the DMS even in the case of a strong spin-orbit (SO) interaction 
in which, naively, only four orbitals are relevant at the top of the 
valence band.\cite{cardona} This model reproduces the Curie temperatures 
observed experimentally for various values of Mn doping and compensations and 
it captures the Curie-Weiss behavior of the magnetization curves in the low 
compensation regime. It also predicts that for Mn doped GaAs the Curie 
temperature would pass through a maximum value $T_C\approx 220K$ for 
$x\approx 12\%$ suggesting that higher Mn doping is not the route to achieve
room temperature $T_C$ with GaAs.

The paper is organized as follows: the general tight binding approach is 
described in Section II; Section III is devoted to systems in which the 
spin-orbit interaction can be neglected; the spin-orbit interaction is 
considered in Section IV by working in a basis formed by total angular 
momentum eigenstates $|j,m_j>$; Section V is devoted to the discussion of 
Coulomb interactions and how to handle the possibility of Mn$^{3+}$ doping. 
Numerical results in 
finite systems for a variety of materials are presented in various subsections
of Section VI; conclusions and a summary appear in Section VII. In the 
Appendix the change of basis matrices are provided.

\section {Tight Binding approach}

Although we will provide a Hamiltonian that can be applied to any 
III-V semiconductor with Zinc-Blende type structure, we will use Mn 
doped GaAs as our example. In GaAs, the Ga and As atoms bond covalently.
Both atoms share the electrons that they have in their 4s and 4p shells. 
Ga shares 3 electrons while As shares 5. The hybridized orbitals have 
character $sp^3$. Although the $s$ and $p$ orbitals have to be considered in 
order 
to obtain the correct band structure, we are interested in light hole doping 
for which we will focus only on the valence band, since to 
facilitate numerical simulations we need to minimize the number of degrees 
of freedom. It is 
well known that around the $\Gamma$ point the valence band of GaAs has 
$j={3\over{2}}$ character, which arises from the original $p$ 
orbitals.\cite{cardona}  Thus, in order to 
construct the simplest model that captures this feature we will  
consider only the $p$ orbitals. The three $p$ orbitals $p_x$, $p_y$ 
and $p_z$ 
in each ion  can be populated with particles with spin up or down.

To study GaAs we should consider two interpenetrating  fcc lattices 
separated by a distance $(a_0/4,a_0/4,a_0/4)$ were $a_0$ is the underlying 
cubic lattice parameter. The Ga ions seat in one of the 
fcc lattices and the As ions in the other. Each ion has 4 nearest 
neighbors of the opposite species located at $(a_0/4,a_0/4,a_0/4)$, 
$(a_0/4,-a_0/4,-a_0/4)$, $(-a_0/4,a_0/4,-a_0/4)$ and $(-a_0/4,-a_0/4,a_0/4)$. 

Since we are only interested in obtaining the valence band we are going to 
consider the bonding combinations of the Ga and As $p$ 
orbitals.\cite{cardona,harrison,chang} 
This leads to an effective fcc lattice with three $p$ bonding orbitals at each 
site that can be occupied by particles with spin up or down. Thus, working in 
this $|p,\alpha>$ basis there are 6 states per site of the fcc lattice.
We will consider the nearest neighbor hopping of these 
particles to construct the effective tight-binding Hamiltonian. The twelve
nearest neighbors are located at $(\pm a_0/2,\pm a_0/2,0)$, 
$(\pm a_0/2,0,\pm a_0/2)$, $(0,\pm a_0/2,\pm a_0/2)$ considering the four sign 
combinations for the three sets of points provided. 

In order to calculate the hoppings we
follow Slater and Koster.\cite{slater} The nearest neighbors in our 
effective fcc lattice are the second nearest neighbors in the original 
diamond structure. From Table I in Ref.\onlinecite{slater}
we see that the relevant overlap integrals in this case are
$$
E_{xx}=l^2(pp\sigma)+(1-l^2)(pp\pi),
$$
$$
E_{xy}=lm[(pp\sigma)-(pp\pi)],
$$
\noindent and
$$
E_{xz}=ln[(pp\sigma)-(pp\pi)].
\eqno(1)
$$

The 12 nearest neighbors in the fcc lattice are labeled $(p,q,r)$ following 
Ref.\onlinecite{slater}. For this geometry, 
two of the indices taking the value $\pm 1$ and the remaining one is 0. 
Since $l={p\over{\sqrt{p^2+q^2+r^2}}}$
with $(p,q,r)=(1,1,0)$, etc. (we are following Slater's 
notation), it follows that $l$, $m$ and $n$ are equal to 0 or $\pm 1/\sqrt{2}$.
Then the hoppings to the twelve neighbors that we will label by
$(\mu,\nu)$ with $\mu$ and $\nu$ taking the values $\pm x$, $\pm y$,and 
$\pm z$, are:

$$
-t_{\rm aa}^{\mu\nu}=E_{xx}(\mu,\nu)={1\over{2}}[(pp\sigma)+(pp\pi)]=
-t_{xx}^{\parallel},
\eqno(2a)
$$
\noindent if either $\mu={\rm a}$ or $\nu={\rm a}$ or
$$
-t_{\rm aa}^{\mu\nu}=E_{xx}(\mu,\nu)=(pp\pi)=
-t_{xx}^{\perp},
\eqno(2b)
$$
\noindent if neither $\nu$ nor $\mu$ are equal to ${\rm a}$, and
$$
-t_{\rm ab}^{\mu\nu}=E_{xy}(\mu,\nu)=\pm {1\over{2}}[(pp\sigma)-(pp\pi)]
=\mp t_{xy},
\eqno(2c)
$$
\noindent with the minus (plus) sign for the case in which $\mu$ and $\nu$ 
have the same (opposite) sign. Also notice that the interorbital hopping is 
only possible when $(\mu,\nu)$ and ab are in the same plane, i.e., there 
is no perpendicular interorbital hopping.

\section {The model without spin-orbit interaction}

If the doped Mn ions go into the parent compound as Mn$^{2+}$ replacing
Ga$^{3+}$, it means that a hole is doped into the $p$ hybridized orbitals
which form the valence band of the undoped semiconductor. This allows us to
write a Hamiltonian that takes into account the nearest neighbor hopping of 
the holes in the $p$ orbitals using the hoppings calculated above, and their
magnetic interaction with the localized randomly doped  Mn$^{2+}$ ions:
$$
{\rm H={1\over{2}}
\sum_{{\bf i,\mu,\nu},\alpha,a,b}t_{ab}^{\bf\mu\nu}
(c^{\dagger}_{{\bf i},\alpha,a}c_{{\bf i+\mu+\nu},\alpha,b}+h.c.)}
$$
$$
+{\rm J\sum_{{\bf I},a}{\bf s_I}^a\cdot{\bf S_I}},
\eqno(3)
$$
\noindent where ${\rm c^{\dagger}_{{\bf i},\alpha,a} }$ creates a hole
at site ${\bf i}=({\rm i_x,i_y,i_z})$ in orbital ${\rm a}$ with spin 
projection $\alpha$,  
${\bf s_I}^{\rm a}$=$\rm \sum_{\alpha\beta} 
c^{\dagger}_{{\bf I},\alpha,a}{\bf{\sigma}}_{\alpha\beta}c_{{\bf
I},\beta,a}$ is the spin of the mobile hole, the  Pauli
matrices are denoted by ${\bf\sigma}$,
${\bf S_I}$ is the localized Mn$^{2+}$
spin $5/2$ at site ${\bf I}$ (covering only a small fraction of the total 
number of 
sites $N$ since Mn replaces a small number of Ga).
$ t_{\rm ab}^{\bf\mu\nu}$ are the hopping amplitudes for the 
holes that were defined in Section II, and
${\rm J>0}$ is an antiferromagnetic (AF) coupling between the spins of
the mobile and localized degrees of freedom.\cite{oka}
The density $\rm \langle n \rangle$ of 
itinerant holes is controlled by a chemical potential. The sites
{\bf i} belong to an fcc lattice and the versors {$\bf\mu\nu$} 
indicate 
the 12 nearest neighbors of each site {\bf i} by taking the values $\pm\hat x$,
$\pm\hat y$, and $\pm\hat z$, with $\mu\neq\nu$. 

Notice that we have only three different hoppings: 
two intraorbital ones $t_{\rm aa}^{\perp}$ and
$t_{\rm aa}^{\parallel}$ and the
interorbital ones $t_{\rm ab}$. All
have the 
same absolute value (but not always 
the same sign) for all combination of orbitals and neighbors. The interorbital 
hoppings that have the sign reversed are the ones towards sites labeled by 
$\mu$ and $\nu$ with opposite signs, i.e. $-t_{\rm xx}^{x,y}=
t_{\rm xx}^{-x,y}$. Also notice that the interorbital 
hoppings occur only in the planes defined by the two orbitals, i.e., they 
vanish in the direction perpendicular to the plane were the two orbitals are
$t_{\rm xz}^{x,y}=0$. This can be seen in the expressions provided in Eq.(1).

In order to 
obtain the material specific values of the hopping parameters we will
 write the Hamiltonian matrix in momentum space for the undoped case, i.e.,
Eq.(3) with $J=0$.
We can use Table II or III in Ref.\onlinecite{slater} to do this task.
The result is given by,
$$
T=
\left(\begin{array}{ccc}
T_x&
-t_{xy}s_{xy}  &
-t_{xy}s_{xz} \\
-t_{xy}s_{xy} &
T_y&
-t_{xy}s_{yz}\\
-t_{xy}s_{xz}&
-t_{xy}s_{yz} &
T_z
\end{array} \right),
\eqno(4)
$$
\noindent for spin up and an identical block for spin down. Here
$T_x=4t_{xx}^{\parallel}(c_xc_y+c_xc_z)+4t_{xx}^{\perp}c_yc_z$,
$T_y=4t_{xx}^{\parallel}(c_xc_y+c_yc_z)+4t_{xx}^{\perp}c_xc_z$,
$T_z=4t_{xx}^{\parallel}(c_xc_z+c_yc_z)+4t_{xx}^{\perp}c_xc_y$, and
$s_{ij}=4s_is_j$ with
$c_i={\rm cos}(ak_i)$, $s_i={\rm sin}(ak_i)$, where in Slater's notation 
$a=a_0/2$, with $a_0$ the Zinc-Blende 
lattice constant and $k_i$ are the momentum components along $i=x, y$, or $z$. 

We mentioned in the introduction that Luttinger and Kohn studied the movement 
of holes in the valence band of semiconductors with the diamond structure. 
Working in momentum space they found an expression for the
Hamiltonian matrix that describes the top of the valence band, i.e., 
the neighborhood of the $\Gamma$ point. In the $|p,\alpha>$
basis the matrix has the form:
\begin{widetext}
$$
\left(\begin{array}{ccc}
Ak_x^2+B(k_y^2+k_z^2) & Ck_xk_y               &       Ck_xk_z \\
Ck_xk_y               & Ak_y^2+B(k_x^2+k_z^2) &       Ck_yk_z \\
Ck_xk_z               & Ck_yk_z               & Ak_z^2+B(k_x^2+k_y^2)
\end{array} \right).
\eqno(5)
$$
\end{widetext}
There is an identical block for spin down. $A$, $B$ and $C$ are constants that 
can be defined in terms of the
Luttinger parameters $\gamma_1$, $\gamma_2$, and $\gamma_3$(\onlinecite{Lut}) 
which are material specific.
The accepted values for GaAs are
$(\gamma_1,\gamma_2,\gamma_3)=(6.85,2.1,2.9)$.\cite{cardona,macdonald} 
The constants are given by:
$$
A=-{\hbar^2\over{2m}}(\gamma_1+4\gamma_2),
$$
$$
B=-{\hbar^2\over{2m}}(\gamma_1-2\gamma_2),
$$
$$
C=-6{\hbar^2\over{2m}}\gamma_3,
\eqno(6)
$$
\noindent where $m$ is the mass of the bare electron.

Remembering that 
Eq.(5) is an approximation valid for the top of the valence band (i.e. the 
$\Gamma$ point), we can expand
the cosines and sines in Eq.(4) and we obtain the matrix shown in Eq.(5) if we 
disregard a constant term along the diagonal that just shifts the top of 
the valence band to $8t_{xx}^{\parallel}+4t_{xx}^{\perp}$ instead of 0.
Comparing the coefficients we obtain
expressions for the hoppings in terms of the Luttinger parameters and the 
lattice constant $a_0=2a$:\cite{chang1}
$$
t_{xx}^{\parallel}={\hbar^2\over{8ma^2}}(\gamma_1+4\gamma_2)=
{\hbar^2\over{2ma_0^2}}(\gamma_1+4\gamma_2),
$$
$$
t_{xx}^{\perp}={\hbar^2\over{8ma^2}} (\gamma_1-8\gamma_2)=
{\hbar^2\over{2ma_0^2}} (\gamma_1-8\gamma_2),
$$
$$
t_{xy}={3\hbar^2\over{4ma^2}}\gamma_3={3\hbar^2\over{ma_0^2}}\gamma_3.
\eqno(7)
$$
Since our goal is to write a tight-binding Hamiltonian for holes that will 
dope 
at most the bottom of the band, we will reverse the signs of the hoppings 
since the 
band obtained in Eq.(5) gets reflected with respect to zero by reversing the 
signs of $A$, $B$, and $C$.\cite{foot} Then, for GaAs, $a_0=5.64\AA$,
\cite{cardona} and from Eq.(7) we obtain:
$$
t_{xx}^{\parallel}=-1.82eV,\quad
t_{xx}^{\perp}=1.20eV,\quad
t_{xy}=-2.08eV.
\eqno(8)
$$

%

It is important to notice that the evaluation of the hopping 
parameters given by Eq.(7) requires three independent parameters to fix their 
values. However, if we look at the expression for the hoppings given in 
Eq.(2) in 
terms of overlap integrals it would appear incorrectly as if only two 
parameters were 
necessary and the three hoppings should be interrelated. The evaluation Eq.(7)
is more accurate though because it considers, in a phenomenological
way, the influence of the neglected bands in the shape of the valence band
at $\Gamma$.

\section {Spin-Orbit interaction}

The spin-orbit interaction is important in GaAs and in other 
III-V semiconductors in which the ion V has a large mass. It mixes the angular 
momentum ($l=1$ for the $p$ orbitals) with the holes' spin degrees of freedom 
producing states with $j={3\over{2}}$ and $j={1\over{2}}$. In a cubic lattice 
with the diamond symmetry, Luttinger and Kohn showed that at the $\Gamma$ 
point the states with $j={1\over{2}}$ get separated from 
those with $j={3\over{2}}$ which are the relevant states at the top of the 
valence band. As a result, naively only four states per site, instead of six, 
become relevant when the spin-orbit interaction is considered and the doping 
is light.\cite{KL} However, as it will be shown in Section VI, at the levels 
of doping of interest ($x\geq 3$\%) the heavy and light hole states that 
become populated 
are sufficiently separated from the $\Gamma$ point that the $j=1/2$ 
contribution becomes important.

To take into account the spin-orbit interaction we will have to perform a 
change 
of basis from $|p,\alpha>$ to $|j,m_j>$. This change of basis has been studied 
by Kohn and Luttinger.\cite{KL} 

The Luttinger-Kohn matrix (Eq.5) in the $|j,m_j>$ basis is presented in 
Eq.(A8) of Ref.\onlinecite{macdonald}. 

The six states that form the basis are:
$$
|1>=|{3\over{2}},{3\over{2}}>, \quad
|2>=|{3\over{2}},-{1\over{2}}>, \quad
|3>=|{3\over{2}},{1\over{2}}>,
$$
$$
|4>=|{3\over{2}},-{3\over{2}}>, \quad
|5>=|{1\over{2}},{1\over{2}}>, \quad
|6>=|{1\over{2}},-{1\over{2}}>.
\eqno(9)
$$
Applying the same change of basis to 
$U=\left(\begin{array}{cc}T&0\\0&T\end{array}\right)$ with $T$ given in Eq.(4)
we obtain the $6\times 6$ matrix $U'=MUM^{-1}$ where $M$ is the change of 
basis matrix provided in Appendix I.

\begin{widetext}                           
$$
U'= \left(\begin{array}{cccccc}
H_{hh} & -c & -b & 0 &{b\over{\sqrt{2}}}&c\sqrt{2}\\
-c^*   & H_{lh} &  0&b&-b^*\sqrt{3\over{2}}&-d\\
-b^*& 0 & H_{lh} & -c&d&-b\sqrt{3\over{2}} \\
0 & b^* & -c^* & H_{hh}&-c^*\sqrt{2}&{b^*\over{\sqrt{2}}}\\
{b^*\over{\sqrt{2}}}&-b\sqrt{3\over{2}}&d^*&-c\sqrt{2}&H_{so}&0\\
c^*\sqrt{2}&-d^*&-b^*\sqrt{3\over{2}}&{b\over{\sqrt{2}}}&0&H_{so}
\end{array} \right),
\eqno(10)
$$
\end{widetext}
\noindent with
$$
H_{hh}=4t_{x,x}^{\parallel}c_xc_y+
2(t_{x,x}^{\perp}+t_{x,x}^{\parallel})(c_xc_z+c_yc_z),
$$
$$
H_{lh}={2\over{3}}(5t_{x,x}^{\parallel}+t_{x,x}^{\perp})(c_yc_z+c_xc_z)
+{4\over{3}}(t_{x,x}^{\parallel}+2t_{x,x}^{\perp})c_xc_y,
$$
$$
b={-4\over{\sqrt{3}}}t_{x,y}(s_x s_z+i s_y s_z),
$$
$$
c={2\over{\sqrt{3}}}(t_{x,x}^{\parallel}-t_{x,x}^{\perp})(c_xc_z-c_yc_z)
-{4i\over{\sqrt{3}}}
t_{x,y}s_xs_y,
$$
$$
d=4({\sqrt{2}\over{3}}(t_{x,x}^{\parallel}-t_{x,x}^{\perp})c_xc_y-
{(t_{x,x}^{\parallel}-t_{x,x}^{\perp})\over{3\sqrt{2}}}(c_xc_z+c_yc_z)),
$$
$$
H_{so}=4{(2t_{x,x}^{\parallel}+t_{x,x}^{\perp})\over{3}}
(c_xc_y+c_xc_z+c_yc_z)+\Delta_{SO},
\eqno(11)
$$
\noindent where $\Delta_{SO}$ is the spin-orbit splitting which is tabulated 
for different materials.\cite{cardona} The matrix in Eq.(10) agrees 
with Eq.A8 in Ref.\onlinecite{macdonald} in the limit of small $k$.

\subsection{Hoppings between $|j,m_j>$ States in Real Space}

Now we need to calculate the hoppings in real space between the 6 orbitals
characterized by $j={3\over{2}}$ and $m_j=\pm {3\over{2}}$ and 
$\pm {1\over{2}}$ and $j={1\over{2}}$ with $m_j=\pm{\tilde 1\over{2}}$.

To do this, we need to express the $c$ operators that appear in Eq.(3) in terms
of the $c$ operators in the basis $|j,m_j>$ with the help of the basis 
transformation given in the Appendix:

\begin{widetext}
$$
\left(\begin{array}{cccccc}
c^{\dagger}_{{\bf i},{3\over{2}},{3\over{2}}}&
c^{\dagger}_{{\bf i},{3\over{2}},-{1\over{2}}}&
c^{\dagger}_{{\bf i},{3\over{2}},{1\over{2}}}&
c^{\dagger}_{{\bf i},{3\over{2}},-{3\over{2}}}&
c^{\dagger}_{{\bf i},{1\over{2}},{1\over{2}}}&
c^{\dagger}_{{\bf i},{1\over{2}},-{1\over{2}}}
\end{array} \right) M=
\left(\begin{array}{cccccc}
c^{\dagger}_{{\bf i},\uparrow,x}&c^{\dagger}_{{\bf i},\uparrow,y}&
c^{\dagger}_{{\bf i},\uparrow,z}&
c^{\dagger}_{{\bf i},\downarrow,x}&
c^{\dagger}_{{\bf i},\downarrow,y}&c^{\dagger}_{{\bf i},\downarrow,z}
\end{array} \right)
\eqno(12)
$$
\end{widetext}
\noindent and
$$
M^{-1}\left(\begin{array}{c}
c_{{\bf i},{3\over{2}},{3\over{2}}}\\
c_{{\bf i},{3\over{2}},-{1\over{2}}}\\
c_{{\bf i},{3\over{2}},{1\over{2}}}\\
c_{{\bf i},{3\over{2}},-{3\over{2}}}\\
c_{{\bf i},{1\over{2}},{1\over{2}}}\\
c_{{\bf i},{1\over{2}},-{1\over{2}}}
\end{array} \right)=
\left(\begin{array}{c}
c_{{\bf i},\uparrow,x}\\
c_{{\bf i},\uparrow,y}\\
c_{{\bf i},\uparrow,z}\\
c_{{\bf i},\downarrow,x}\\
c_{{\bf i},\downarrow,y}\\
c_{{\bf i},\downarrow,z}
\end{array} \right).
\eqno(13)
$$

We obtain:
$$
c^{\dagger}_{{\bf i},\sigma,x}=\sigma({-1\over{\sqrt{2}}}
c^{\dagger}_{{\bf i},{3\over{2}},\sigma {3\over{2}}}+{1\over{\sqrt{6}}}
c^{\dagger}_{{\bf i},{3\over{2}},-\sigma {1\over{2}}})-{1\over{\sqrt{3}}}
c^{\dagger}_{{\bf i},{1\over{2}},-\sigma{1\over{2}}},
$$
$$
c^{\dagger}_{{\bf i},\sigma,y}=-i({1\over{\sqrt{2}}}
c^{\dagger}_{{\bf i},{3\over{2}},\sigma {3\over{2}}}+{1\over{\sqrt{6}}}
c^{\dagger}_{{\bf i},{3\over{2}},-\sigma {1\over{2}}}-{\sigma\over{\sqrt{3}}}
c^{\dagger}_{{\bf i},{1\over{2}},-\sigma{1\over{2}}}),
$$
$$
c^{\dagger}_{{\bf i},\sigma,z}=\sqrt{{2\over{3}}}
c^{\dagger}_{{\bf i},{3\over{2}},\sigma {1\over{2}}}-\sigma\sqrt{{1\over{3}}}
c^{\dagger}_{{\bf i},{1\over{2}},\sigma {1\over{2}}},
\eqno(14a)
$$
\noindent and
$$
c_{{\bf i},\sigma,x}=\sigma({-1\over{\sqrt{2}}}
c_{{\bf i},{3\over{2}},\sigma {3\over{2}}}+{1\over{\sqrt{6}}}
c_{{\bf i},{3\over{2}},-\sigma {1\over{2}}})-
{1\over{\sqrt{3}}}
c_{{\bf i},{1\over{2}},-\sigma {1\over{2}}},
$$
$$
c_{{\bf i},\sigma,y}=i({1\over{\sqrt{2}}}
c_{{\bf i},{3\over{2}},\sigma {3\over{2}}}+{1\over{\sqrt{6}}}
c_{{\bf i},{3\over{2}},-\sigma {1\over{2}}})-
{i\sigma\over{\sqrt{3}}}
c_{{\bf i},{1\over{2}},-\sigma {1\over{2}}},
$$
$$
c_{{\bf i},\sigma,z}=\sqrt{{2\over{3}}}
c_{{\bf i},{3\over{2}},\sigma {1\over{2}}}-\sigma
{1\over{\sqrt{3}}}
c_{{\bf i},{1\over{2}},\sigma {1\over{2}}}.
\eqno(14b)
$$
Replacing in Eq.(3) and rearranging the terms we find that the intraorbital
hoppings are given by:

$$
t^{x,y}_{\sigma{3\over{2}},\sigma{3\over{2}}}=t_{xx}^{\parallel}, \quad
t^{x,y}_{\sigma{1\over{2}},\sigma{1\over{2}}}={t_{xx}^{\parallel}+
2t_{xx}^{\perp}\over{3}}
$$
$$
t^{y,z}_{\sigma{3\over{2}},\sigma{3\over{2}}}=t^{x,z}_{\sigma{3\over{2}},
\sigma{3\over{2}}}={(t_{xx}^{\parallel}+t_{xx}^{\perp})\over{2}}
$$
$$
t^{y,z}_{\sigma{1\over{2}},\sigma{1\over{2}}}=t^{x,z}_{\sigma{1\over{2}},
\sigma{1\over{2}}}={(5t_{xx}^{\parallel}+
t_{xx}^{\perp})\over{6}}
$$

$$
t^{\mu,\nu}_{\sigma{\tilde 1\over{2}},\sigma{\tilde 1\over{2}}}=
{2t_{xx}^{\parallel}+
t_{xx}^{\perp}\over{3}}
\eqno(15a)
$$

Now let us consider the inter-orbital hoppings between orbitals with $j=3/2$:

$$
t^{x,z}_{\rm \sigma a,-\sigma a'}=-t^{y,z}_{\rm\sigma a,-\sigma a'}
={(t_{xx}^{\perp}-t_{xx}^{\parallel})\over{\sqrt{12}}},
$$
$$
t^{x,y}_{\rm a,-a'}=(t^{x,y}_{\rm a,-a'})^*=t^{y,z}_{\sigma 1/2,
\sigma 3/2}=(t^{y,z}_{\sigma 3/2,\sigma 1/2})^*=
{i\over{\sqrt{3}}}t_{xy}^{\mu,\nu}
$$
$$
t^{x,z}_{\rm\sigma a,\sigma a'}=\sigma{t_{xy}^{\mu,\nu}
\over{\sqrt{3}}}, \quad
t^{x,y}_{\rm\sigma a,\sigma a'}=0.
$$
\noindent and
$$
t^{\mu,\nu}_{\rm\sigma a,-\sigma a}=0,
\eqno(15b)
$$
\noindent where ${\rm a\neq a'}$ in the above expressions.

There are no interorbital hoppings between the two $j=1/2$ orbitals, i.e,
$$
t^{\mu,\nu}_{\rm\sigma a,-\sigma a}=0,
\eqno(15c)
$$
\noindent if $a={\tilde 1\over{2}}$, i.e., the symbol that we use
to represent $m_j$ for the orbitals with $j={1\over{2}}$. 

The interorbital hoppings between $j=3/2$ and $j=1/2$ orbitals are given by:
$$
t^{xz}_{\rm\sigma{\tilde 1\over{2}},\sigma{3\over{2}}}=
t^{xz}_{\rm\sigma{3\over{2}},\sigma{\tilde 1\over{2}}}=
-{t_{xy}^{\mu,\nu}\over{\sqrt{6}}}
$$
$$
t^{yz}_{\rm\sigma{3\over{2}},\sigma{\tilde 1\over{2}}}=
(t^{yz}_{\rm\sigma{\tilde 1\over{2}},\sigma{3\over{2}}})^*=
-i{t_{xy}^{\mu,\nu}\over{\sqrt{6}}}
$$
$$
t^{xz}_{\rm\sigma{1\over{2}},-\sigma{\tilde 1\over{2}}}=
t^{xz}_{\rm\sigma{\tilde 1\over{2}},-\sigma{1\over{2}}}=
{t_{xy}^{\mu,\nu}\over{\sqrt{2}}}
$$
$$
t^{yz}_{\rm\sigma{1\over{2}},-\sigma{\tilde 1\over{2}}}=
t^{yz}_{\rm\sigma{\tilde 1\over{2}},-\sigma{1\over{2}}}=
i \sigma {t_{xy}^{\mu,\nu}\over{\sqrt{2}}}.
$$
$$
t^{x,y}_{\rm\sigma{1\over{2}},\sigma{\tilde 1\over{2}}}=
t^{x,y}_{\rm\sigma{\tilde 1\over{2}},\sigma{1\over{2}}}
=-\sigma{\sqrt{2}(t_{xx}^{\perp}-t_{xx}^{\parallel})\over{3}},
$$
$$
t^{x,z}_{\rm\sigma{1\over{2}},\sigma{\tilde 1\over{2}}}=
t^{y,z}_{\rm\sigma{1\over{2}},\sigma{\tilde 1\over{2}}}=
t^{x,z}_{\rm\sigma{\tilde 1\over{2}},\sigma{1\over{2}}}=
t^{y,z}_{\rm\sigma{\tilde 1\over{2}},\sigma{1\over{2}}}
=\sigma{(t_{xx}^{\perp}-t_{xx}^{\parallel})\over{3\sqrt{2}}},
\eqno(15d)
$$

The actual numerical values are obtained using Eq.(7). \cite{chang2} 
Notice that the hoppings can also be read from the matrix elements in Eq.(10).


\subsection{Hund Coupling and the Classical Limit of the Localized Spin}

The Hund term in Eq.(3) has the form
$$
H_{\bf I}=J{\bf S_I\cdot s_I},
\eqno(16)
$$
\noindent where ${\bf S_I}$ is a dimensionless quantum spin 5/2 operator,
\cite{spin,Linnar} 
${\bf s_I}$ is a dimensionless quantum spin 1/2 operator, and $J$, 
which has units 
of energy (eV), is determined experimentally.\cite{oka,matsu,macdonald}

In order to perform numerical calculations avoiding the sign problem\cite{sign}
which 
arises when four-fermion interactions, such as in Eq.(16), are decoupled, we
are going to take the classical limit for the localized spins. This is a good
approximation given the large value, 5/2, of this spin. The classical limit
is given by $\hbar\rightarrow 0$, $S\rightarrow\infty$. In order to take the 
limit correctly Eq.(16) is rewritten as
$$
H_{\bf I}=J[S(S+1)]^{1/2}{\cal S}_{\bf I}\cdot{\bf s_I},
\eqno(17)
$$
\noindent where ${\cal S}_{\bf I}$ is a unit spin quantum operator.

It is important to 
realize that the Hund coupling $J$ is proportional to $\hbar$ since the spin 
operators in Eq.(16) are dimensionless. This means that $\lim_{\hbar\to 0}J=0$.
However, $\lim_{\hbar\to 0,S\to\infty}J[S(S+1)]^{1/2}=
\lim_{\hbar\to 0,S\to\infty}JS$, should be finite. Defining
$$
\lim_{\hbar\to 0,S\to\infty}JS=J_c,
\eqno(18)
$$
\noindent in the classical limit Eq.(17) becomes
$$
H_{\bf I}=J_c{\cal S}_{\bf I}\cdot{\bf s_I},
\eqno(19)
$$
\noindent where now
$$
{\cal S}_{\bf I}=({\rm sin\theta_{\bf i}cos\phi_{\bf i},sin\theta_{\bf i}
sin\phi_{\bf i},cos\theta_{\bf i}}),
\eqno(20)
$$
\noindent is a classical unit vector. ${\bf s_I}$ is still a quantum spin-1/2
operator and $J_c$ is a parameter. The value of $J_c$ should be determined by 
comparing numerical results to experimental data. For Mn doped GaAs 
comparing numerical values of the Curie temperature to experimental data, 
we have observed \cite{we} that $J_c\approx J$. This indicates that
$$
\lim_{\hbar\to 0,S\to\infty}JS=J,
\eqno(21)
$$
\noindent which is equivalent to 
$$
\lim_{\hbar\to 0,S\to\infty}\hbar S=\hbar_0=6.58 \times 10^{-16} eV,
\eqno(22)
$$
\noindent i.e., the classical limit according to 
Ref.\onlinecite{classical}. 

Additional support for these results is obtained measuring the splitting of the
4 degenerate states at the top of the valence band as a function of $J$. 
Experimentalists use this split to determine $J$.\cite{spin,Linnar,Szc} 
According to mean-field calculations the splitted levels have energy $\pm B_G$ 
and $\pm 3B_G$ with\cite{Furd,Dietl,Szc}
$$
B_G={1\over{6}}JxS,
\eqno(23)
$$
\noindent with $S=5/2$.\cite{spin,Linnar} According to Eq.(23) 
$$
B_G^c=\lim_{\hbar\to 0,S\to\infty}B_G={1\over{6}}J_cx\equiv{1\over{6}}Jx.
\eqno(24)
$$
Our numerical results indicate that the split of the top of the valence 
band depends linearly on $J$ for values up to $J\leq 0.5$ eV 
(see subsection VI.C).
In Fig.1 we show the mean field value of $B_G$, i.e., $B_G^{MF}$, calculated 
using Eq.(23) as a function of $J$ for $x=0.085$ and $x=0.05$ indicated with 
doted lines. We also display the extrapolated linear behavior obtained 
numerically for $J\leq 0.5$ eV and
the same values of $x$ at $T\to 0$ with our Hamiltonian. The curves indicating
$B_G^C$ in the figure have been obtained by monitoring the energy of the 
four lowest eigenvalues of Eq.(31) as a function of $J$. 
In the linear regime we observe that the ratio between $B_G^{MF}$ given by 
Eq.(23) and the measured numerical value in the classical approximation 
$B_G^C$ is equal to $S=2.5$ as expected. 
The excellent numerical agreement 
between $B_G^{MF}/S$ and $B_G^c$ supporting the result of Eq.(24) is clear in
Fig.1.

\begin{figure}[thbp]
\begin{center}
\includegraphics[width=8cm,clip,angle=0]{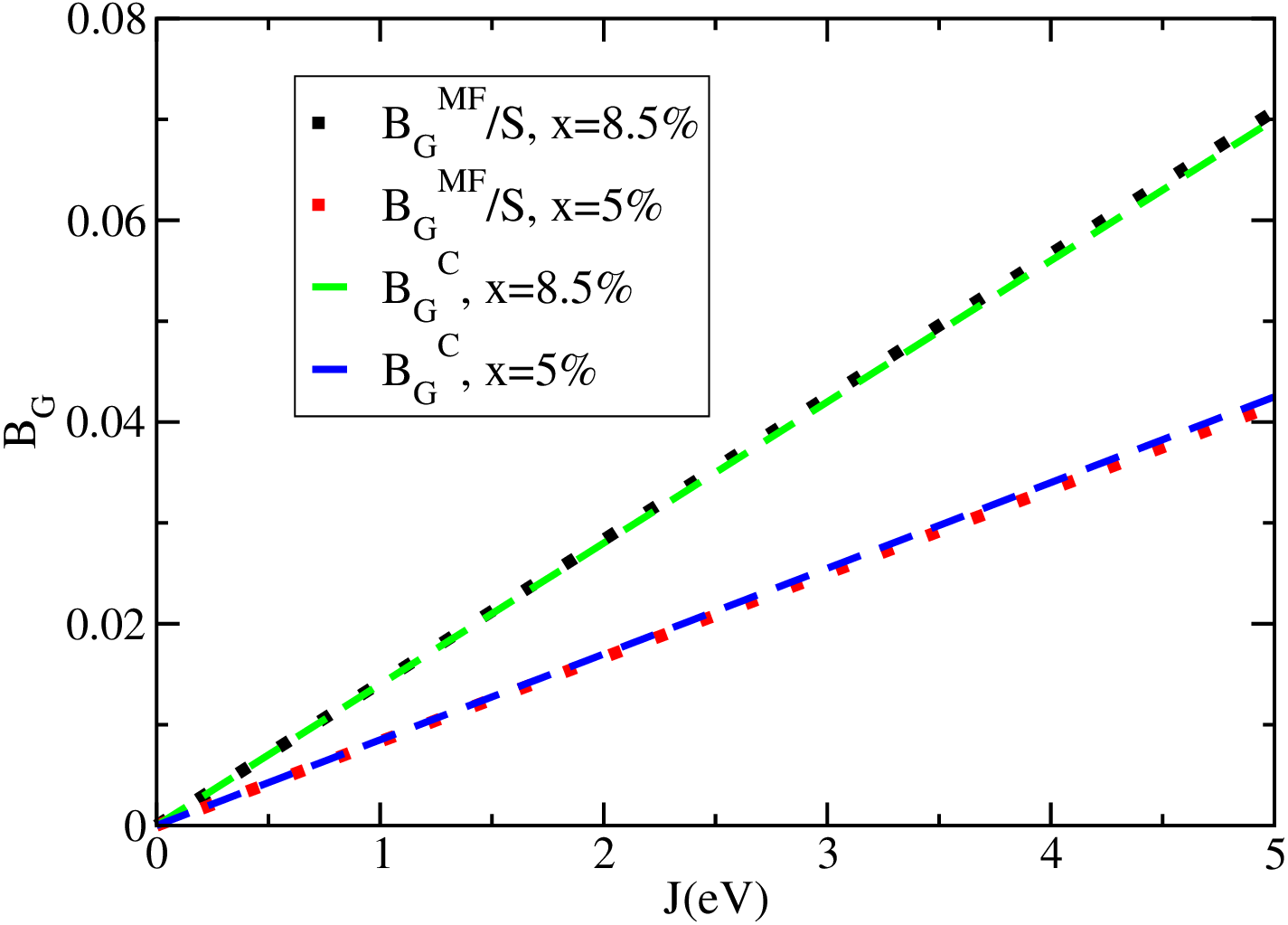}
\vskip 0.3cm
\caption{(color online) $B_G^{MF}/S$ vs $J$ for $x=0.085$ and 0.05 
obtained using Eq.(23) 
(doted lines) and $B_G^c$ vs $J$ calculated numerically 
(extrapolating from the linear regime) 
with our model for the same values of $x$ (dashed lines).}
\label{gaas}
\end{center}
\end{figure}

Now let us consider the Hund coupling term in Eq.(3) in 
the $|j,m_j>$ basis. Eq.(A.10) in Ref.\onlinecite{macdonald} 
has expressions for the spin 
operators in the $|j,m_j>$ basis: 

$$
s_x=  \left(\begin{array}{cccccc}
0 & 0 & {1\over{2\sqrt{3}}} & 0 &{1\over{\sqrt{6}}}&0\\
0 & 0 & {1\over{3}}& {1\over{2\sqrt{3}}}&-{1\over{3\sqrt{2}}}&0 \\
{1\over{2\sqrt{3}}} & {1\over{3}}&0&0&0&{1\over{3\sqrt{2}}}\\
0&{1\over{2\sqrt{3}}} &0&0&0&-{1\over{\sqrt{6}}}\\
{1\over{\sqrt{6}}}&-{1\over{3\sqrt{2}}}&0&0&0&-{1\over{6}}\\
0&0&{1\over{3\sqrt{2}}}&-{1\over{\sqrt{6}}}&{-1\over{6}}&0
\end{array} \right)
\eqno(25a)
$$
$$
s_y= i \left(\begin{array}{cccccc}
0 & 0 & {-1\over{2\sqrt{3}}} & 0 &-{1\over{\sqrt{6}}}&0\\
0 & 0 & {1\over{3}}& {-1\over{2\sqrt{3}}}&-{1\over{3\sqrt{2}}}&0 \\
{1\over{2\sqrt{3}}} & {-1\over{3}}&0&0&0&-{1\over{3\sqrt{2}}}\\
0&{1\over{2\sqrt{3}}} &0&0&0&-{1\over{\sqrt{6}}}\\
{1\over{\sqrt{6}}}&{1\over{3\sqrt{2}}}&0&0&0&{1\over{6}}\\
0&0&{1\over{3\sqrt{2}}}&{1\over{\sqrt{6}}}&-{1\over{6}}&0
\end{array} \right)
\eqno(25b)
$$
$$
s_z=  \left(\begin{array}{cccccc}
{1\over{2}} & 0 & 0 & 0&0&0 \\
0 & {-1\over{6}} & 0& 0&0&-{\sqrt{2}\over{3}} \\
0 & 0&{1\over{6}}&0&-{\sqrt{2}\over{3}}&0\\
0&0 &0&{-1\over{2}}&0&0\\
0&0&-{\sqrt{2}\over{3}}&0&-{1\over{6}}&0\\
0&-{\sqrt{2}\over{3}}&0&0&0&{1\over{6}}
\end{array} \right)
\eqno(25c)
$$

Then
$$
J{\bf S_I\cdot s_I}\rightarrow
J(S_{\bf I}^xs_{\bf I}^x+S_{\bf I}^ys_{\bf I}^y+S_{\bf I}^zs_{\bf I}^z)
\eqno(26)
$$

\noindent where 
$$
s_{\bf I}^{\alpha}={\bf c_I}^{\dagger} s_{\alpha}{\bf c_I}
\eqno(27)
$$
\noindent with
$$
{\bf c_I}^{\dagger}=\left(\begin{array}{cccccc}
c^{\dagger}_{{\bf i},{3\over{2}}}&
c^{\dagger}_{{\bf i},-{1\over{2}}}&
c^{\dagger}_{{\bf i},{1\over{2}}}&
c^{\dagger}_{{\bf i},-{3\over{2}}}&
c^{\dagger}_{{\bf i},{\tilde 1\over{2}}}&
c^{\dagger}_{{\bf i},-{\tilde 1\over{2}}}
\end{array} \right)
\eqno(28)
$$
\noindent and under the classical approximation that we just discussed,

$$
S_{\bf I}^x={\rm sin\theta_{\bf i}cos\phi_{\bf i}}, \quad
S_{\bf I}^y={\rm sin\theta_{\bf i}sin\phi_{\bf i}}, \quad
S_{\bf I}^z={\rm cos\theta_{\bf i}}.
\eqno(29)
$$
Notice that the Mn ions replace Ga so 
they will be present in a subset of the points of the fcc lattice that we 
are considering.
The values of $J$ can be obtained from comparing experimental data with 
theoretical models and are material dependent. 
In the notation of Ref.\onlinecite{Dietl},
$J$ in Eq.(26) is given by $\beta N_0$ where $\beta$ has units of $eVnm^3$ 
and $N_0$ is the concentration of cation sites given by 
$4a_0^{-3}$ where 
$a_0$ is the lattice parameter of the material. $\beta$ is assumed to 
depend only on the characteristics of the parent material and it is considered
the same for all III-V semiconductors. Notice that $\beta$ is called $J_{pd}$ 
by some authors.\cite{macdonald} Ref.\onlinecite{Dietl} estimates $J$ for 
GaN and other materials assuming that the accepted value for GaAs is accurate 
and given by 
$J=4 \beta a_0^{-3}=-1.2eV$.\cite{oka} However, in the literature the
value of $J$ for GaAs ranges between 
$-0.89eV\geq J\geq -3.34eV$.\cite{oka,matsu,macdonald}
Using the parameters for GaAs we can estimate that 
$-0.04eVnm^3\geq\beta_{III-V}\geq -0.15eVnm^3$.

Then, $J$ for a general III-V material $M$ is given by
$$
J_{M}={4\beta_{III-V}\over{a_{0_{M}}^3}}.
\eqno(30)
$$

The sign depends on the definition. It seems to be antiferromagnetic for GaAs 
which means that we need to take it as a positive number in our Hamiltonian.
The sign of $J$ is still under discussion for some of the III-V compounds such 
as GaN.\cite{GaNsign} However, under the classical spin approximation 
discussed above, the sign of $J$ in the Hamiltonian becomes irrelevant since 
the eigenvalues of $H$ are the same for both signs.  

Notice that using the parameters provided in this paper, numerical 
calculations of the critical temperature will be obtained directly in eV.

\subsection{Six-Orbital Real-Space Hamiltonian}

Combining the results in the previous subsections the Hamiltonian in 
the $|j,m_j>$ basis can be written as:
$$
{\rm H={1\over{2}}
\sum_{{\bf i,\mu,\nu},\alpha,\alpha',a,b}(t_{\alpha a,\alpha' b}^{\bf\mu\nu}
c^{\dagger}_{{\bf i},\alpha a}c_{{\bf i+\mu+\nu},\alpha' b}+h.c.)}
$$
$$
+\Delta_{SO}\sum_{{\bf i},\alpha}c^{\dagger}_{{\bf i},\alpha{\tilde1\over{2}}}
c_{{\bf i},\alpha{\tilde1\over{2}}}+
{\rm J\sum_{{\bf I}}{\bf s_I}\cdot{\bf S_I}},
\eqno(31)
$$
\noindent where ${\rm a,b}$ take the values ${1\over{2}}$, ${\tilde1\over{2}}$ 
or ${3\over{2}}$, 
$\alpha$ and $\alpha'$ can be $1$ or $-1$, and the Hund term 
is given by Eq.(26). Notice that {$\bf\mu+\nu$} are the twelve 
vectors indicating the twelve nearest 
neighbor sites of each ion at site {\bf i} and 
that {\bf I} are random sites in the fcc lattice. 
To reproduce the correct physics
we will have to work with an almost filled valence band, i.e., a very low 
density of holes. Eq.(31) is the Hamiltonian that has to be programmed in the 
computer simulations. 
Since there are six states per site, a $6N_s\times 6N_s$ 
matrix needs to be diagonalized at each step of a Monte Carlo (MC) simulation; 
$N_s$ is the number of sites in the fcc lattice.

The material dependent hopping parameters are given by Eq.(7) and (15), and 
the Hund coupling by Eq.(30) (or from experiments if available). 
$\Delta_{SO}$ is given by the magnitude of the gap between
the top of the valence band and the split-off band induced by the spin-orbit
interaction. Its value is 0.341eV for GaAs, 0.75eV for GaSb, and 0.017eV for
GaN.\cite{cardona} Thus, it is clear that all the parameters in the proposed 
Hamiltonian are fixed, i.e., {\it there are no free parameters}.

\section{Additional Effects of the doped Mn Ions}

In this section we will consider, at a phenomenological 
level, the effects of Mn doping that go beyond the mere introduction of a 
localized spin
at the doping site $I$ and the addition of holes into the system. 

Up to this point we have implicitly assumed that the Mn $d$ levels are deep 
into the valence band of the parent compound so that when a Mn
ion replaces Ga in GaAs, Ga$^{3+}$ is replaced by Mn$^{2+}$ and a hole is 
introduced in the $p$ orbitals. In this case, negative charge 
with respect to the background at the Mn site
creates an attractive Coulomb potential for the hole which, in the case of one
single Mn ion, contributes to producing 
a bound state at 0.1eV above the top of the valence 
band.\cite{matsu,Linnar} This bound state is expected to generate an 
impurity band at least for very light amounts of doping. This impurity band, 
due to 
hybridization, eventually will merge with the valence band as the amount 
of doping increases. For GaAs it is 
believed that the merging occurs for $x<0.01$\cite{Fleurov,Popescu} and for 
this reason the Coulomb interaction is not expected to play a role in the 
relevant doping regime. However, the explicit addition of  a Coulomb term to 
the Hamiltonian is important in order to reproduce the GaAs behavior at 
very low dopings and, in materials for which the hole binding energy is larger,
such as GaP and GaN.\cite{Koro,Kreis} Thus, below, in subsection A, we are 
going to describe how to introduce Coulomb attraction in our model.
 
Another effect that is in general ignored in studies of (Ga,Mn)As is the
presence of Mn$^{3+}$ ions. While experiments appear to indicate that 
Ga$^{3+}$ is replaced by Mn$^{2+}$
in GaAs,\cite{Schneider,spin} Mn$^{3+}$ has been reported
in GaP\cite{Kreis} and GaN.\cite{Kro,Graf} Thus, in order to extend the 
present model to other materials, in subsection B we will describe a 
phenomenological way of considering the Mn $d$ orbitals. 

\subsection{Coulomb Potential}

The Coulomb potential between the localized impurities and the doped holes
is neglected in most models for DMS\cite{Gonz,Mat,Ken,Sarma,Schl,Xu}
since the magnetic interaction appears to be sufficient to capture 
qualitatively many properties of these materials including ferromagnetism, 
and because it is believed that at the levels of doping for which the material 
is metallic it will not 
play a relevant role.\cite{Fleurov,Popescu} However, these assumptions need to 
be confirmed by actual calculations. Very few attempts to include the Coulomb
interaction studying its effects with unbiased approaches have been carried 
out. 
The case of a single Mn impurity in GaAs was studied in Ref.\onlinecite{guill}.
They considered the long range Coulomb potential suplemented by a central 
cell correction with a square-well shape that is routinely applied in 
calculations of bound state energies for impurities in 
semiconductors.\cite{pante} This is the portion of the potential that is 
expected to be relevant at higher doping since the long-range part will be 
screened. This effect has been incorporated by many 
authors\cite{taka,calde,hwang,Popescu} by considering an on-site central-cell 
potential of the form:
$$
H_C=-V\sum_{I}n_I.
\eqno(32)
$$
The value of $V$ varies widely in the literature. Many authors determine it by 
finding the value that added to $J$ reproduces the energy of the single hole
bound state.\cite{guill,taka,Popescu} However, the value of $V$ depends 
strongly on 
the characteristic of the model used such as the total bandwidth, the number 
of orbitals, etc. In other investigations $V$ is considered a free parameter 
that can 
even take negative values, i.e., repulsive rather than attractive according to 
the notation of Eq.(32).\cite{calde,hwang} Repulsive potentials are sometimes 
introduced to reproduce apparent $x$ dependences of the magnetic coupling $J$ 
observed in some experiments on Mn doped GaN\cite{DIETL07} and 
CdS.\cite{TWOR94} 

However, it was pointed out in Ref.\onlinecite{Popescu} that an on-site range
potential prevents the overlap of localized hole-wave functions that 
should occur with increasing doping. It was proposed that a nearest-neighbor 
range potential should be used, and MC calculations performed on a highly 
doped 
single orbital model showed that an impurity band generated by the on-site
potential vanished when a nearest-neighbor range potential of the same strength
was used. In the fcc lattice considered here, the extended potential has the 
form
$$
H_C^{nn}=-V\sum_{I}({1\over{N_I}}n_I+\sum_{\mu,\nu}
{1\over{N_{I+\mu+\nu}}}n_{I+\mu+\nu}),
\eqno(33)
$$
\noindent where $V$ is added at the site I in which 
the impurity is located and 
at its 12 nearest neighbors I+$\mu+\nu$. $N_I$ indicates
the number of impurity sites that surround the impurity site $I$. Calculations 
using both Eq.(32) and (33) will be presented in Section VI.E.

\subsection{Mn$^{3+}$}

The phenomenology described in the previous subsection is expected to occur
when the Mn $d$ orbitals lie deep into the valence band so that Mn$^{2+}$ 
replaces the III$^{3+}$ ion and the doped hole goes into the $p$ orbitals. 
However, if the 
Mn $d$ levels were in the parent compound's gap, the 
III$^{3+}$ ion could be replaced by Mn$^{3+}$ with the hole in
its $3d$ shell. The $3d$ shell of Mn is splitted into three $t_{2g}$ states
and two $e_g$ states due to the crystal field. It is believed that the hole
goes into the $t_{2g}$ levels.\cite{hole} Although no Mn$^{3+}$ has been 
observed in GaAs, there are some indications of it in GaP and 
GaN.\cite{Kreis,Kro,Graf}

Ideally, at least the three $t_{2g}$ orbitals in each Mn should be considered
but we would not be able to study the problem numerically due to the large 
number of additional degrees of freedom. Thus, we will follow a more
phenomenological approach by introducing only one extra orbital, which will 
add two extra degrees of freedom per Mn due to the spin of the hole.
For this purpose,  
we extend the six orbital model described by 
Eq.(31) by adding an extra orbital at the impurity sites that can be populated 
by holes with spin up or down. To simplify the calculations we will
consider an extra s-like orbital. Thus, a contribution $H_{p-d}$ given by
$$
H_{p-d}=(V_d-\mu)\sum_{I}n^d_{\bf I}+\sum_{\langle I,J \rangle,\sigma}
(t_{ss}d^{\dagger}_{I,\sigma}d_{J,\sigma}+h.c.)+
$$
$$
\sum_{\langle I,j \rangle,\sigma,\alpha}
(t_{sp}d^{\dagger}_{I,\sigma}c_{j,\alpha}+h.c.)+J\sum_{I,\gamma,\delta}
d^{\dagger}_{I,\gamma}\vec\sigma_{\gamma,\delta}d_{I,\delta}
\cdot{\bf S_I}
\eqno(34)
$$
\noindent will be added to Eq.(31). In Eq.(34), $V_d$ is the parameter 
that determines the position of the Mn $d$ level in relation to the top of the 
valence band. Its value could, eventually, be obtained from experiments
but here, in this first effort, it is going to be treated as a free parameter; 
$\mu$ is the chemical potential that should be added to Eq.(31) to fix the 
number of holes; $n^d_{I}=
\sum_{\sigma}d^{\dagger}_{I,\sigma}d_{I,\sigma}$ is the number operator for 
the holes in the $d$-orbital; $d^{\dagger}_{I,\sigma}$ $(d_{I,\sigma})$ creates
(destroys) a hole with spin $\sigma$ in the (single) ``$d$''-orbital at 
impurity site $I$; $\langle I,J \rangle$ indicate nearest-neighbor impurity 
sites; 
$t_{ss}$ is the phenomenological hopping between holes in the $d$-orbitals of 
nearest-neighbor Mn impurities (notice that this hopping is active only in 
the disorder configurations in which there are Mn ions next to each other).
$t_{sp}$ is the hopping between
holes in the $d$-orbitals at the impurity sites and any of the six $p$-orbitals
in the nearest-neighbor sites (notice that the hoppings between $p$ levels
between nearest neighbor and impurity sites are already contained in Eq.(31)).
Finally, the Hund interaction between the localized spin and the spin of the
hole in the $d$-orbital at the impurity site is considered. If we were not 
working in the classical limit for the spins, a spin two operator 
(instead of S=5/2) will have to be used in Eqs.(34) and (31).

Notice that the $t_{sp}$ hoppings are functions of the direction and, if the 
spin-orbit interaction is considered, of the quantum numbers $j,m_j$. Following
the procedure described in Section II we know that in an fcc lattice
$$
E_{sx}=l(sp\sigma), \quad
E_{sy}=m(sp\sigma),\quad
E_{sz}=n(sp\sigma).
\eqno(35)
$$
\noindent Then the hoppings to the 12 neighbors are:
$$
-t^{\mu,\nu}_{sa}=E_{sa}(\mu,\nu)=(-1)^{\mu+\nu}{(sp\sigma)\over{\sqrt{2}}}=
\mp t_{sp},
\eqno(36)
$$
\noindent if either $\mu=a$ or $\nu=a$, or zero otherwise. Notice that the 
minus (plus) sign in the hopping corresponds to the case in which $\mu$ and
$\nu$ have the same (opposite) sign.

Applying the basis transformation given by Eqs.(12) and (13) to the 
$p$-orbitals
the explicit form of the hoppings between the $d$-orbital and the $(j,m_j)$
orbitals are obtained:

$$
t^{x,y}_{d\uparrow,{3\over{2}}}=-t_{sp}{(1+i)\over{\sqrt{2}}},\quad
t^{x,z}_{d\uparrow,{3\over{2}}}=-t_{sp}{i\over{\sqrt{2}}},
$$
$$
t^{y,z}_{d\uparrow,{3\over{2}}}=-t_{sp}{1\over{\sqrt{2}}},\quad
t^{x,y}_{d\uparrow,-{1\over{2}}}=-t_{sp}{(1-i)\over{\sqrt{6}}},
$$
$$
t^{x,z}_{d\uparrow,-{1\over{2}}}=t_{sp}{i\over{\sqrt{6}}},\quad
t^{y,z}_{d\uparrow,-{1\over{2}}}=-t_{sp}{1\over{\sqrt{6}}},
$$
$$
t^{x,y}_{d\uparrow,{1\over{2}}}=0,\quad
t^{x,z}_{d\uparrow,{1\over{2}}}=t^{y,z}_{d\uparrow,{1\over{2}}}=
it_{sp}{\sqrt{2\over{3}}},
$$
$$
t^{x,y}_{d\uparrow,-{3\over{2}}}=t^{x,z}_{d\uparrow,-{3\over{2}}}=
t^{y,z}_{d\uparrow,-{3\over{2}}}=
t^{x,y}_{d\uparrow,{\tilde 1\over{2}}}=0
$$
$$
t^{x,z}_{d\uparrow,{\tilde 1\over{2}}}=t^{y,z}_{d\uparrow,{\tilde 1\over{2}}}=
-it_{sp}{\sqrt{1\over{3}}},\quad
t^{x,y}_{d\uparrow,-{\tilde 1\over{2}}}=-t_{sp}{(1+i)\over{\sqrt{3}}},
$$
$$
t^{x,z}_{d\uparrow,-{\tilde 1\over{2}}}=-t_{sp}{i\over{\sqrt{3}}},\quad
t^{y,z}_{d\uparrow,-{\tilde 1\over{2}}}=t_{sp}{1\over{\sqrt{3}}},
$$
$$
t^{x,y}_{d\downarrow,{3\over{2}}}=t^{x,z}_{d\downarrow,{3\over{2}}}=
t^{y,z}_{d\downarrow,{3\over{2}}}=
t^{x,y}_{d\downarrow,-{1\over{2}}}=0,
$$
$$
t^{x,z}_{d\downarrow,-{1\over{2}}}=t^{y,z}_{d\downarrow,-{1\over{2}}}=
it_{sp}{\sqrt{2\over{3}}},\quad
t^{x,y}_{d\downarrow,{1\over{2}}}=-t_{sp}{(1+i)\over{\sqrt{6}}},
$$
$$
t^{x,z}_{d\downarrow,{1\over{2}}}=-t_{sp}{i\over{\sqrt{6}}},\quad
t^{y,z}_{d\downarrow,{1\over{2}}}=-t_{sp}{1\over{\sqrt{6}}},
$$
$$
t^{x,y}_{d\downarrow,-{3\over{2}}}=-t_{sp}{(1-i)\over{\sqrt{2}}},\quad
t^{x,z}_{d\downarrow,-{3\over{2}}}=t_{sp}{i\over{\sqrt{2}}},
$$
$$
t^{y,z}_{d\downarrow,-{3\over{2}}}=-t_{sp}{1\over{\sqrt{2}}},\quad
t^{x,y}_{d\downarrow,{\tilde 1\over{2}}}=t_{sp}{(1-i)\over{\sqrt{3}}},
$$
$$
t^{x,z}_{d\downarrow,{\tilde 1\over{2}}}=-t_{sp}{i\over{\sqrt{3}}},\quad
t^{y,z}_{d\downarrow,{\tilde 1\over{2}}}=t_{sp}{1\over{\sqrt{3}}},
$$
$$
t^{x,y}_{d\downarrow,-{\tilde 1\over{2}}}=0, \quad
t^{x,z}_{d\downarrow,-{\tilde 1\over{2}}}=
t^{y,z}_{d\downarrow,-{\tilde 1\over{2}}}= it_{sp}{\sqrt{1\over{3}}}.
\eqno(37)
$$

\section{Numerical results in finite systems}

In this section, we address the results obtained from the study of Eq.(31) in 
a finite fcc lattice. In reciprocal space, the allowed momentum 
values form a cubic structure with $\delta k_i=2\pi/(a_0N_i)$ where $i=x,y$, or
$z$ and $N_i$ is the number of unit cubic cells in the sample along the $i$ 
direction. Since in laboratory samples $N_i$ is of the order of Avogadro's 
number we can replace $k_i$ by a continuous variable. However, this is not 
true in the small finite systems that can be studied numerically even with the 
most powerful computers currently available. Only when $J=0$ 
we can diagonalize Eq.(10) using continuous values of ${\bf k}$. The results 
for GaAs along high symmetry directions in the first Brillouin zone (FBZ) 
are indicated by the continuous line in Fig.2.   
\begin{figure}[thbp]
\begin{center}
\includegraphics[width=8cm,clip,angle=0]{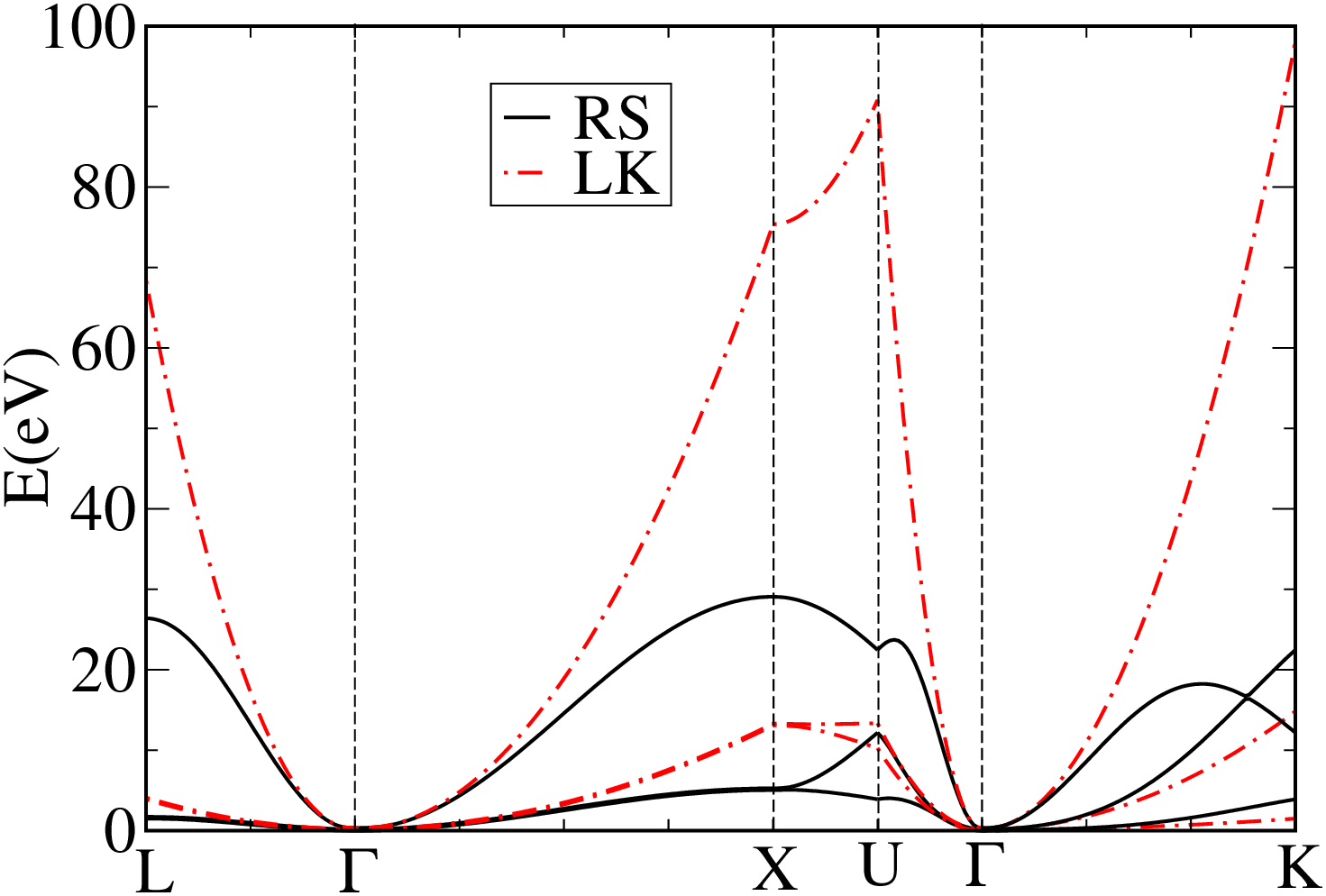}
\vskip 0.3cm
\caption{(color online) Band structure for GaAs obtained diagonalizing 
Eq.(10) 
(continuous lines). The dashed lines indicate the results for the 
Luttinger-Kohn model.}
\label{gaas}
\end{center}
\end{figure}
As expected, the bottom of 
the band is at $\Gamma$ and we observe the heavy hole and light hole bands 
along the high symmetry lines in the Brillouin zone shown in the figures.
Since we are using ``hole'' language the top of the electronic valence band 
appears here as the bottom of the ``hole'' valence band. 
The dashed lines are the eigenvalues of Eq.(A8) in Ref.\onlinecite{macdonald},
i.e., the Luttinger-Kohn model results. In order to check the agreement with 
our results around the point $\Gamma$ we shifted our curves by the value
$8t_{xx}^{\parallel}+4t_{xx}^{\perp}$ so that the bottom of our valence 
band is at zero. 
It can be seen that the agreement between the curves obtained with our tight 
binding model and the Luttinger-Kohn ones is excellent at the $\Gamma$ point
of the valence band. However, our model captures better the band dispersion
away from the center of the Brillouin zone and the 
agreement with ARPES results is remarkable.\cite{ARPES}


When $J\neq 0$ and a finite number of magnetic impurities are considered, 
the diagonalization of Eq.(31) has to be performed in real space. 
In Fig.3 we show the discrete energy eigenvalues along the 
high symmetry lines in the FBZ obtained numerically diagonalizing Eq.(31) using
finite clusters (still with $J=0$). $N=N_x=N_y=N_z$ indicates the number of 
unit cells considered along each spatial direction. 
\begin{figure}[thbp]
\begin{center}
\includegraphics[width=8cm,clip,angle=0]{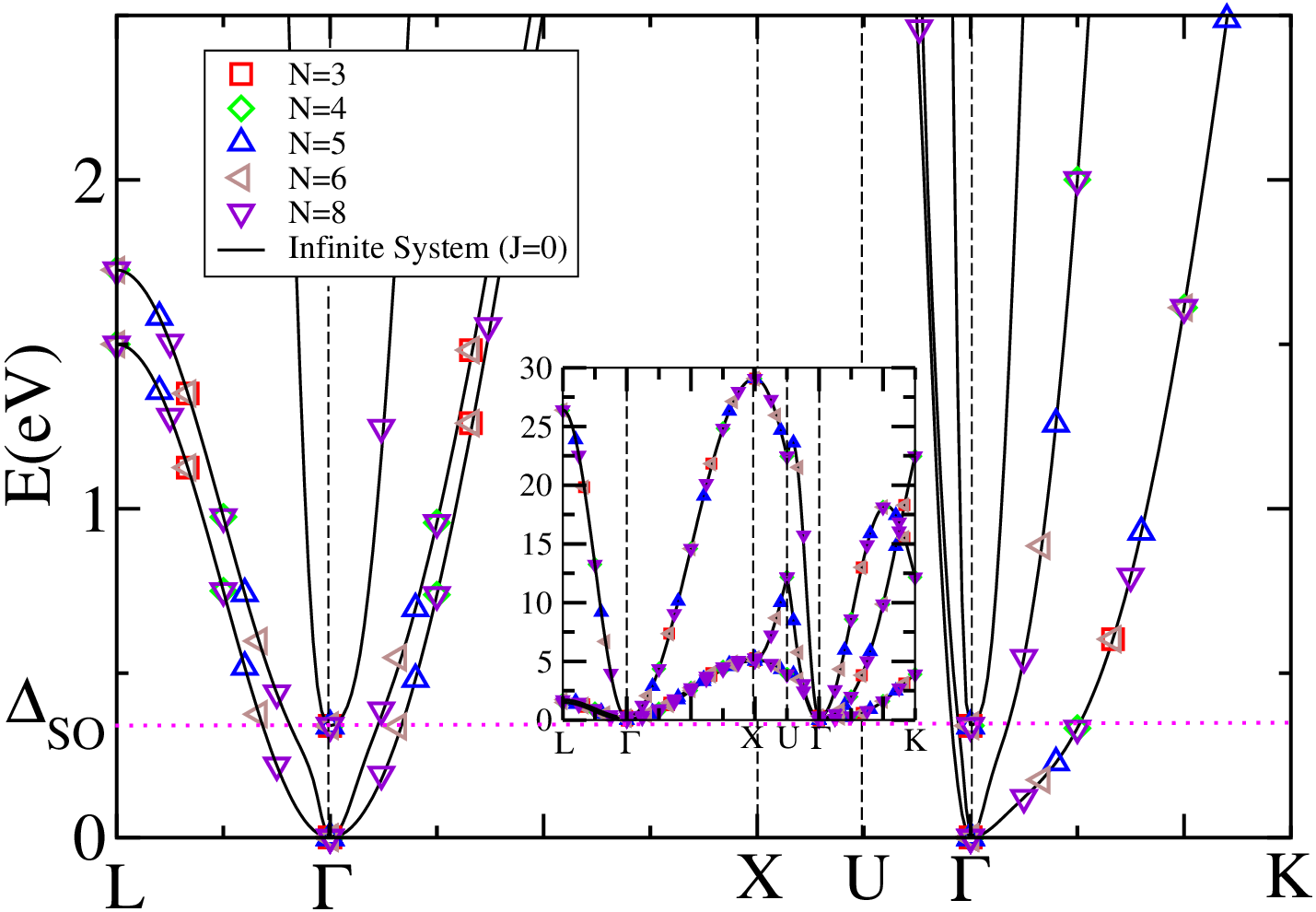}
\vskip 0.3cm
\caption{(color online) Band structure for GaAs obtained diagonalizing 
Eq.(10) 
(black lines). The symbols indicate results obtained using finite lattices 
in real space diagonalizing Eq.(31) with J=0. As the lattice size 
increases, more momentum states in the First Brillouin zone
become available. The number of momentum states available inside the FBZ is
given by $4N^3$.}
\label{gaex}
\end{center}
\end{figure}
Numerically, we study clusters that contain $N$ cubes of side $a_0$ 
along each of the 3 spatial directions.
Since there are 4 ions associated to each site of the cube in an fcc lattice,
the total number of Ga ions in the numerical simulation is given by
$N_{Ga}=4N^3$. This means that there is an equal number of points inside the 
FBZ in momentum space. As already mentioned, the 
discrete lattice in momentum space is cubic. The side of the smallest cube is 
given by $b=2\pi/Na_0$ which
is the size of the mesh corresponding to $N$ cubic cells along each of the 
three spatial directions. The first 
Brillouin zone has the shape of a truncated octahedron defined by well 
known high symmetry points such as $L=(\pi/a_0,\pi/a_0,\pi/a_0)$, 
$X=(2\pi/a_0,0,0)$, $U=(2\pi/a_0,\pi/2a_0,\pi/2a_0)$, 
$K=(3\pi/2a_0,3\pi/2a_0,0)$, etc.\cite{equiv} 

The finiteness of the lattices imposes a constraint on the minimum Mn 
concentrations that can be reached: values of $x$ larger than 0.2\% can be
studied in lattices with $N=4$, i.e. 256 sites. The results presented in this 
work have been obtained using standard Monte Carlo simulations for the 
localized spins configurations which requires, at each iteration, the exact 
diagonalization of a $4N^3N_o\times 4N^3N_o$ fermionic matrix, where $N_o=6$
for Eq.(31) and 8 when the ``$d$-orbital'' is included.\cite{manga} Using 
state-of-the-art computers such as the Cray XT3 supercomputer at ORNL we are 
currently able to perform production runs for $N=4$ and some runs with $N=5$ 
and 6, and also study non cubic lattices, to monitor finite size effects. 
We are working 
on adapting newly developed (approximated) numerical techniques such as the 
TPEM approach\cite{furu} to this problem with the hope of achieving 
production runs for $N=6$ and 8.
 

\subsection{Four Band Approximation}

In the case of III-V semiconductors with strong spin-orbit interaction, i.e.,
large $\Delta_{SO}$ such as GaAs and GaSb, only the four $j=3/2$ orbitals
are often considered to study the top of the valence band.
\cite{macdonald,KL} This approximation would certainly simplify 
numerical simulations since the number of degrees of freedom per lattice site
is reduced from six to four. 


However, we have found an important drawback. As it can be seen in Fig.4 
the band dispersions under the four-orbital approximation (continuous line)
agrees with the dispersions for the six-orbital case (dashed line)
only in a very small region about $\Gamma$ with radius 
$\delta ka_0 \approx 0.02\pi$ while the study of many 
representative dopings such as $x=8.5\%$ with $p=0.7$ involves states at a 
radius $\delta ka_0\approx 0.5\pi$. In this situation the four-orbital 
approximation  no longer holds.
Along the $\Gamma-L$, $\Gamma-X$, and $\Gamma-U$ directions the
heavy-hole dispersion is well reproduced for a large momentum range
but the very close in energy light-hole state is 
not captured by the 4 orbital approximation. Even worse, along the $\Gamma-K$
direction, the flatness characteristic of the heavy-hole band is 
missed in the 4 orbital approach. Thus, the 4 orbital model would only be 
adequate to describe very lightly doped cases (probably in the insulating 
regime) beyond the reach of finite lattice simulations and not very relevant 
for the high $T_C$ regime.
\begin{figure}[thbp]
\begin{center}
\includegraphics[width=8cm,clip,angle=0]{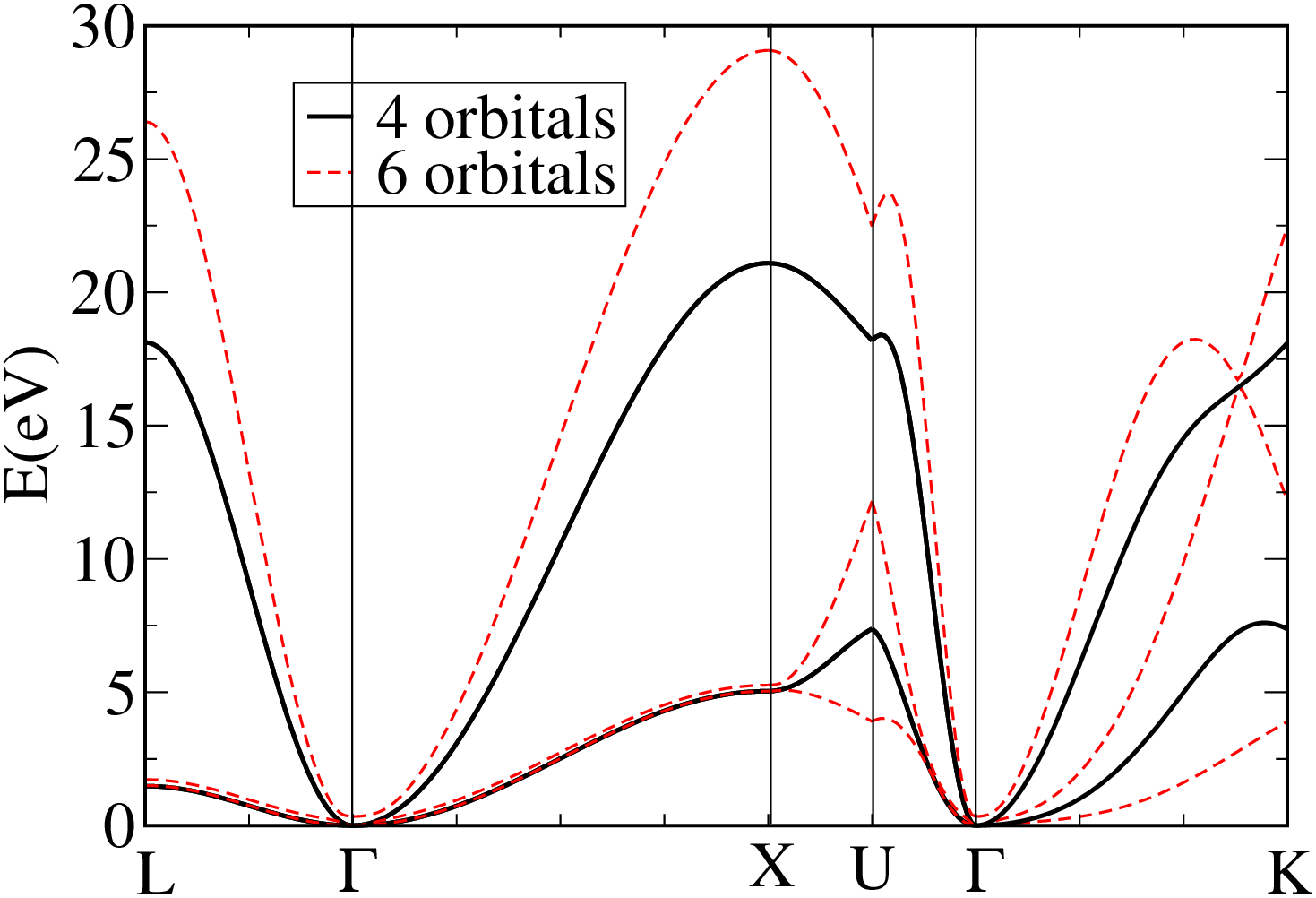}
\vskip 0.3cm
\caption{(color online) Band structure for GaAs obtained diagonalizing 
Eq.(10) 
(dashed lines). The band structure using the 4 orbital approximation is
indicated with continuous lines.}
\label{gaex}
\end{center}
\end{figure}

An example of the shortcomings of the 4 orbital model is the incorrect form of
the magnetization M as a function of temperature (T) that it provides.
In Fig.5a the magnetization curve for $x=8.5\%$ Mn doped GaAs calculated with 
the 4-orbital approximation (circles) is compared with the one given by 
Eq.(31) (diamonds). The Curie-Weiss (CW) shape 
observed in experiments\cite{Pot} is only captured by the 6-orbital 
model\cite{we} while the 4-orbital approximation reproduces the concave shape 
observed in previous numerical efforts dealing mostly with single orbital 
models.\cite{Gonz,Ken} In addition, the value of $T_C$ is underestimated in 
this approach.

\subsection{Spin-Orbit Interaction}

As pointed out in the previous subsection, a problem of early unbiased 
numerical efforts was the failure to reproduce the Curie-Weiss shape of the 
magnetization curves obtained in experiments of Mn doped GaAs with low 
compensation. While this shape was characteristic of most mean-field based 
approaches\cite{dietl00, macdonald}, a concavity consistent with percolative 
behavior was observed otherwise.\cite{Gonz,Ken}  

According to our results the CW behavior is very sensitive to the dispersion
curves and the number of orbitals. 
Numerical simulations of the model presented here have shown that the 
spin-orbit interaction is crucial to obtain the CW shape of the 
magnetization curves.\cite{we} In Fig.5a we present magnetization curves 
obtained in
lattices with 256 Ga ions for $J=1.2eV$, $x=8.5\%$, and the hole density 
$p=0.7$. The curve  with $\Delta_{SO}=0.34eV$, indicated with diamonds, is in 
excellent agreement with 
the experimental data for Mn doped GaAs exhibiting CW behavior. It also provides 
an appropriate value of $T_C$. On the other hand, the curve with 
$\Delta_{SO}=0$ (indicated with squares), 
shows a linear increase of the magnetization below $T_C$.
\begin{figure}[thbp]
\begin{center}
\includegraphics[width=8cm,clip,angle=0]{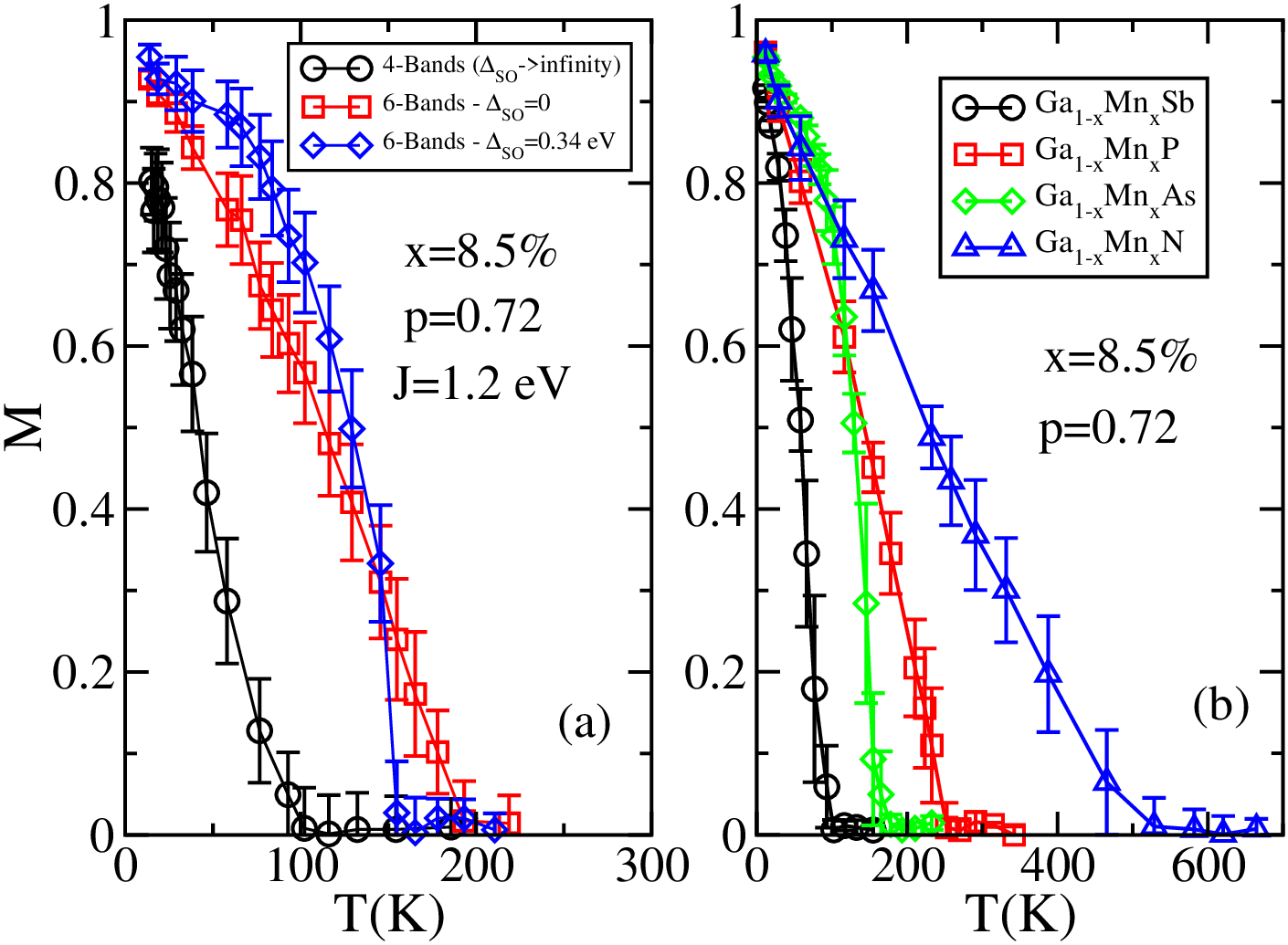}
\vskip 0.3cm
\caption{(color online) Magnetization vs temperature calculated numerically 
with x=8.5\% of 
Mn, and a density of holes p=0.72 for (a) GaAs; the diamonds 
indicate 
the results with spin-orbit interaction while the squares show the points 
for which the spin-orbit interaction has been neglected. Results using 
the 4-orbital approximation, i.e. $\Delta_{SO}\rightarrow\infty$ are 
denoted by the circles. (b) GaSb (circles), GaAs (diamonds), GaP (squares),
and GaN (triangles).
The magnetization is measured as
$\cal M=\sqrt{{\bf M}\cdot{\bf M}}$,
with ${\bf M}$ the vectorial magnetization. As a consequence,
for fully disordered spins, $\cal M$ is
still nonzero due to the ${\bf S_I}^2$=1 contributions, causing a finite
value at large temperatures (${\cal M}(T\to\infty)=1/\sqrt{xN_{Ga}}$)
unrelated to ferromagnetism. Thus, we plotted
${\rm M}=({\cal M}-{\cal M}(T\to\infty))/(1-{\cal M}(T\to\infty))$, i.e. the
background was subtracted.}
\label{gaas}
\end{center}
\end{figure}
We have also used Eq.(31) to calculate the magnetization of other doped III-V
compounds with higher and lower spin-orbit coupling than GaAs. GaSb has a 
stronger spin-orbit coupling $\Delta_{SO}=0.75$eV. Eq.(31) can be used for GaSb
with the hoppings obtained using Eq.(7) and (15) using the Luttinger parameters
$(\gamma_1,\gamma_2,\gamma_3)=(13.3,4.4,5.7)$.\cite{cardona} $J$ is obtained 
using Eq.(30) and, since
$a_0=6.10\AA$\cite{cardona}, $J$=0.96eV. 

The magnetization, denoted by the 
circles in Fig.5b shows CW behavior but $T_C\approx 100$K is lower than for 
GaAs at the same doping and compensation. The result roughly agrees with 
mean-field estimates.\cite{dietl00} We also present results for Mn doped GaP 
which has a much smaller spin-orbit coupling ($\Delta_{SO}=0.08$eV) than GaAs.
The Luttinger parameters are $(4.05,0.49,1.25)$\cite{cardona} and we use 
$J=1.34$eV.
In this case we obtain a magnetization curve with an almost linear temperature
dependence (squares in Fig.5b). This linear behavior has been obtained in 
measurements of (Ga,Mn)P with $1.8\%\le x\le 3.8\%$\cite{Farsh} but it is not 
clear whether it will continue up to 8.5\% doping if this value were reached. 

For completeness, we also present the results for (Ga,Mn)N. A $T_C\approx 500$K,
well above room temperature, is obtained in this case. This result is in 
agreement with mean-field predictions, assuming the doped Mn in the state
Mn$^{2+}$,\cite{dietl00,dietl02,mas} and with some experimental 
measurements.\cite{gan,Koro} The effects of  Mn-$d$-level participation will 
be discussed in subsection E.

\subsection{$J$-Induced Splitting of the Valence Band at the $\Gamma$-Point}

Many studies of DMS are based on reasonable assumptions on the properties of 
the ground state. Some researchers work in the ``high doping'' regime assuming 
that the doped holes are uniformly distributed in the system and effectively 
doped into the valence band of the material. This is the so-called 
valence-band 
scenario\cite{OHN96,dietl00, macdonald,jung,Sam} in which the disorder 
configurations of the Mn ion are disregarded. The other approach, known as the
impurity-band regime,\cite{IB} considers the limit of very low doping in which 
the holes electrically bound to the impurity cores start forming an impurity 
band due to the overlap of the wave functions. In this case the disordered 
positions of the Mn ions 
would play an important role. Although each scenario must be valid in opposite
doping regimes, it is not clear when the crossover between the two occurs 
(since it
is a function of $x$, $J$, and other parameters) and which one better 
describes the 
experiments is controversial. Even for (Ga,Mn)As both descriptions are applied
to the case of $x$=8.5\%.

A way of shedding light on this issue is by studying the evolution as a 
function of $J$, and for different values of $x$, of the four 
degenerate states at
the top of the valence band ($\Gamma$ point) of the parent compound. 
This splitting is due to the interaction of the 
doped holes with the localized spins and it was briefly described when the 
classical limit was discussed in section IV.A. For small values of $J$, 
at values of $x$ in the 
metallic regime according to Mott's criterion, it is expected that the 
mean-field approach (assuming uniform hole distribution) should be a good 
approximation. The calculations indicate\cite{Furd,Dietl,Szc} that the levels
will split following a linear behavior in $Jx$, i.e., $E_0\pm B_G$ and
$E_0\pm 3B_G$ with $B_G$ given in Eq.(23) and (24).  

We have evaluated the eigenvalue distribution for different Mn disorder 
configurations (indicated with different symbols in Fig.6) as a function of
$J$ for $x$=8.5\% (Fig.6a) and 5\% (Fig.6b). For both values of $x$ it can be 
seen 
that the mean-field prediction is accurate for $J\le 0.5$eV; for $J$=1.2eV, 
i.e., the accepted value for GaAs, the eigenvalues have lower energies 
than the mean-field 
ones indicating a stronger distortion of the bands due to the interactions. 
However,
the eigenvalues appear to be independent of the disorder configurations 
indicating that the holes are non-localized; this uniform hole distribution 
continues up to $J\approx 4$eV which is the value at which our numerical 
simulations indicate that an impurity band starts to develop in the density 
of states.\cite{we} For larger values of $J$ there is a large spread in the 
eigenvalue distribution as a function of the disorder. Notice that the effects 
of disorder are much stronger for $x$=5\% than for $x$=8.5\% which 
corroborates the assumption that the role of disorder diminishes as $x$ 
increases.  It is interesting 
to notice that, in disagreement with the mean-field expectation that the 
spread in the eigenvalue distribution should increase with $x$, we actually 
observe that as $x$ increases the 
width of the eigenvalues band decreases.

\begin{figure}[thbp]
\begin{center}
\includegraphics[width=8cm,clip,angle=0]{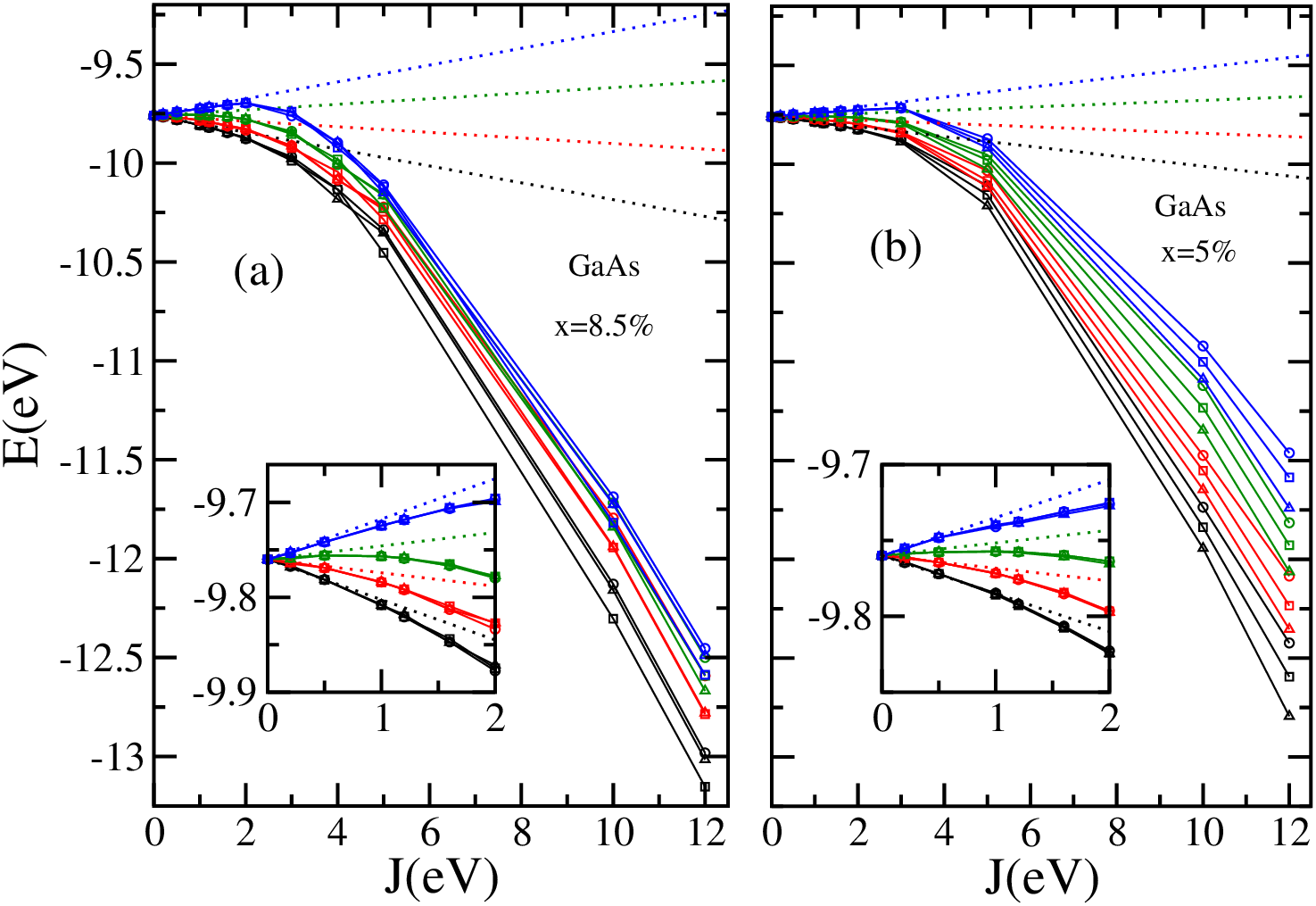}
\vskip 0.3cm
\caption{(color online) Numerically determined evolution of the four 
degenerate 
states at the 
top of the valence as a function of $J$ for (a) $x=0.085$ and (b) for 
$x=0.05$ in a 256 sites lattice. The 
different symbols indicate different Mn disorder configurations. The 
dotted lines correspond to the mean-field predictions. The insets show the 
range in $J$ for which agreement between the numerical and mean-field results 
occur.}
\label{gaas}
\end{center}
\end{figure}

These results indicate that $J=1.2$eV corresponds to an intermediate region 
in which in the metallic regime the holes are not localized (if in the 
metallic regime of $x$)but the mean-field approximation needs corrections.

\subsection{Dependence of $T_C$ with $x$}

An important issue in DMS is the dependence of the Curie temperature with the 
amount of magnetic impurities. Mean-field calculations\cite{macdonald} 
predict that $T_C$ increases linearly with $x$ but numerical simulations in 
simple systems
indicate that $T_C$ reaches a maximum at an optimal value of $x$ ($x_{opt}$)
and decreases afterwards.\cite{Gonz} The optimal value of $x$ was found to be
a function of $J$. For large $J$, $x_{opt} \rightarrow 1$ while 
$x_{opt} \approx 0.25$ 
for the values of $J$ that capture the physics of (Ga,Mn)As. 
Experimentally, in (Ga,Mn)As, 
$T_C$ has been found to increase with $x$ up to about $x$=10\%.\cite{Pot} 
Recently, $x$=12.2\%-
21.3\% has been achieved\cite{tanaka} but the highest $T_C$ observed has been 
170K, i.e., no higher than the record value ($T_C$=173K) obtained 
with $x$=9\%.\cite{wang} In fact, for both annealed and as grown 
samples it was 
observed that $T_C$ reaches a broad maximum for $x \sim$10\%-15\% and 
decreases 
afterwards. It is not clear whether the decrease is due to intrinsic 
properties of the material or to a larger amount of compensation. In Fig.7
we present results for $T_C$ vs $x$ obtained with Eq.(31) using the parameters 
for (Ga,Mn)As. Experimental points are presented for comparison. As $x$ 
increases longer thermalizations are needed in the numerical simulations. For 
$p=0.5$, 0.75 and 1 we have obtained reasonable agreement with the experimental
data for the annealed samples. We obtained a maximum $T_C=(225\pm 20)$K for
$x=12.5\%$ in the non-compensated case, i.e. $p=1$. Thus, according to our 
calculations, even larger Mn doping of GaAs would not allow to reach a Curie
temperature above room temperature.

\begin{figure}[thbp]
\begin{center}
\includegraphics[width=8cm,clip,angle=0]{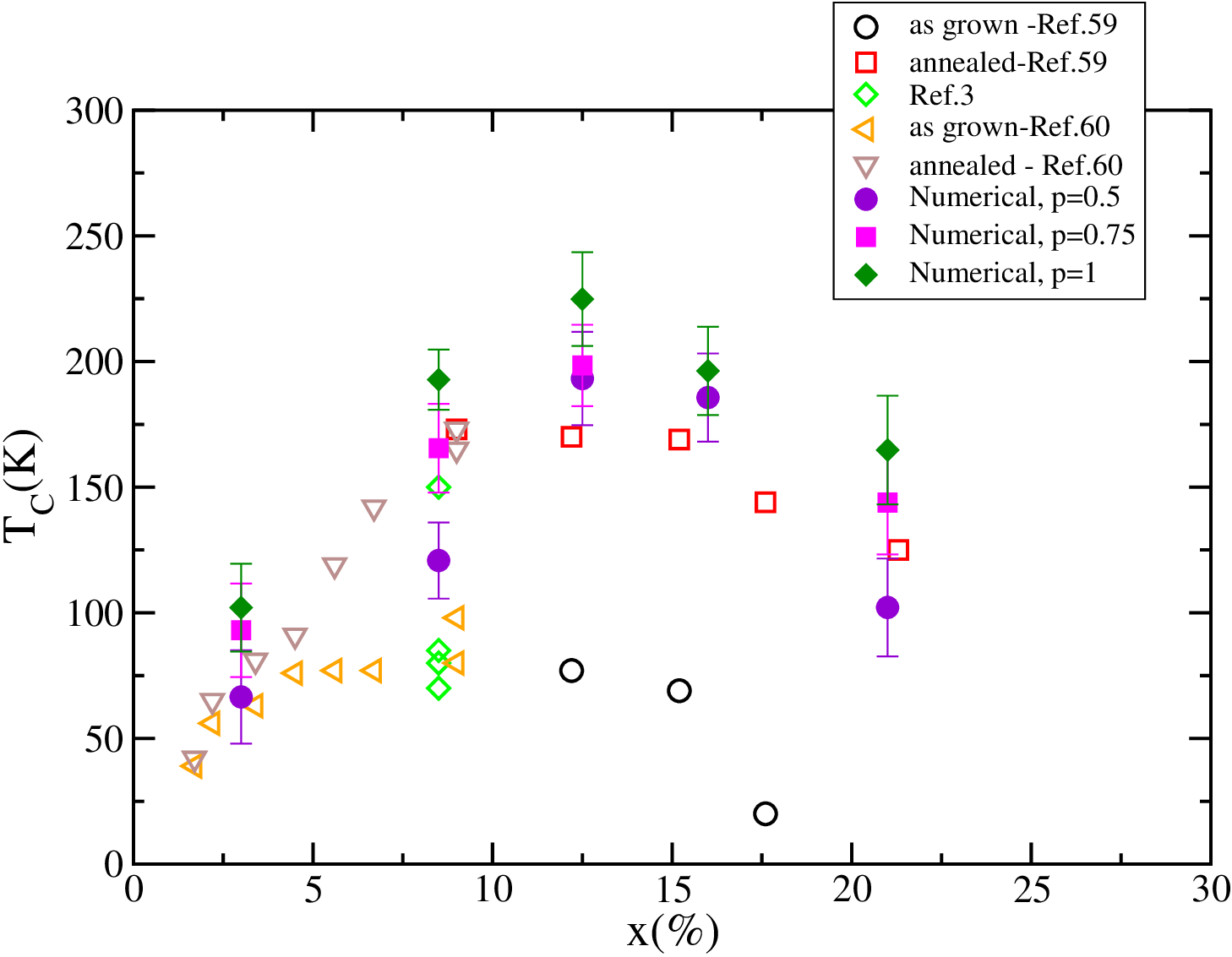}
\vskip 0.3cm
\caption{(color online) Numerically determined $T_C$ as a function of $x$ 
for $\rm{Ga_{\it 1-x}Mn_{\it x}As}$
for different values of compensation $p$. Experimental data are also shown.}
\label{gaas}
\end{center}
\end{figure}

\subsection{Coulomb Attraction at Mn Sites}

Numerical calculations of the density of states (DOS) using Eq.(31) for the 
case of (Ga,Mn)As have indicated that the chemical potential is in the valence
band for dopings x=3\% or higher.\cite{we} As discussed in Section V.A it has 
been argued by some researchers that this outcome may be the result of 
neglecting the effects of Coulomb attraction between the localized Mn$^{2+}$ 
ions and the doped holes. In this subsection we will study the 
effects of adding the terms Eq.(32) and Eq.(33) to Eq.(31). 
In Ref.\onlinecite{Popescu} it was estimated that $V=-1.9eV$ for Mn doped GaAs.
In Fig.8 it can be seen that the DOS for both x=8.5\% (Fig.8a) and 3\% (Fig.8b)
remains basically unchanged when either an onsite or a nearest neighbor 
Coulomb potential of the above magnitude is added. The chemical potential is 
still in the valence band and no hint of impurity band appears. Since even for 
the zero range potential the holes are already delocalized, there is no 
difference when the longer range potential is considered. Similar results (no
IB in the DOS) are obtained for the case of (Ga,Mn)P using 
V=-2.4 eV.\cite{Popescu} This indicates that for the values of the 
parameters for (Ga,Mn)As, the addition of Coulomb attraction is irrelevant in 
the range of doping of interest, i.e., x$\ge 3\%$.

\begin{figure}[thbp]
\begin{center}
\includegraphics[width=8cm,clip,angle=0]{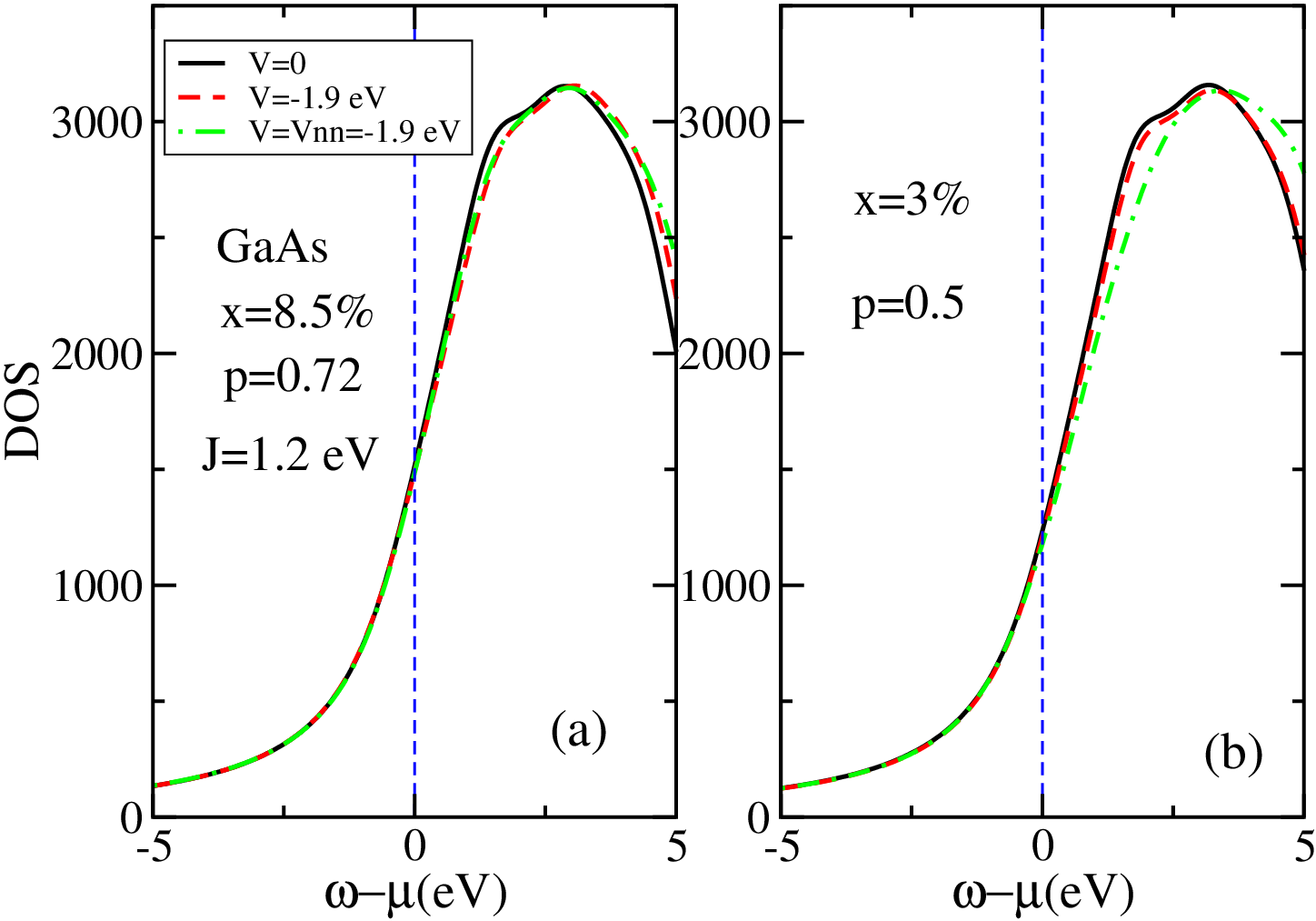}
\vskip 0.3cm
\caption{(color online) Numerically determined density of states for Mn doped 
GaAs with
(a) $x=0.085$ and and $p=0.72$ and (b) for $x=0.05$ and $p=0.5$. The 
different lines indicate different ranges of the Coulomb potential. $Vnn$ 
indicates the value of the Coulomb potential at the impurity's nearest 
neighbor sites.}
\label{gaas}
\end{center}
\end{figure}

However, for larger (but unphysical) values of $J$ our results confirm the 
conclusions 
of Ref.\onlinecite{Popescu} regarding
the importance of considering a nearest-neighbor range potential in order to
accurately obtain the value of $x$ for which the crossover between 
the IB and VB 
description occurs. In Fig.9 we show the DOS for the unphysical case in which
the hopping parameters correspond to GaAs but we set
$J$=7eV instead of the realistic value 1.2eV for x=8.5\% (dashed line 
in Fig.9). We chose $J$=7 eV because it is close to the value 
for 
which a clearly separated IB develops in the DOS.\cite{we} By adding an on-site
Coulomb potential of strength -3.5eV an IB develops (continuous line) and the 
chemical potential is located there. Clearly the holes are trapped at the Mn 
sites 
due to the large potential. However, if a nearest-neighbor range potential with
the same strength is applied, the holes become delocalized and the IB band in 
the DOS disappears (dot-dashed line).

\begin{figure}[thbp]
\begin{center}
\includegraphics[width=8cm,clip,angle=0]{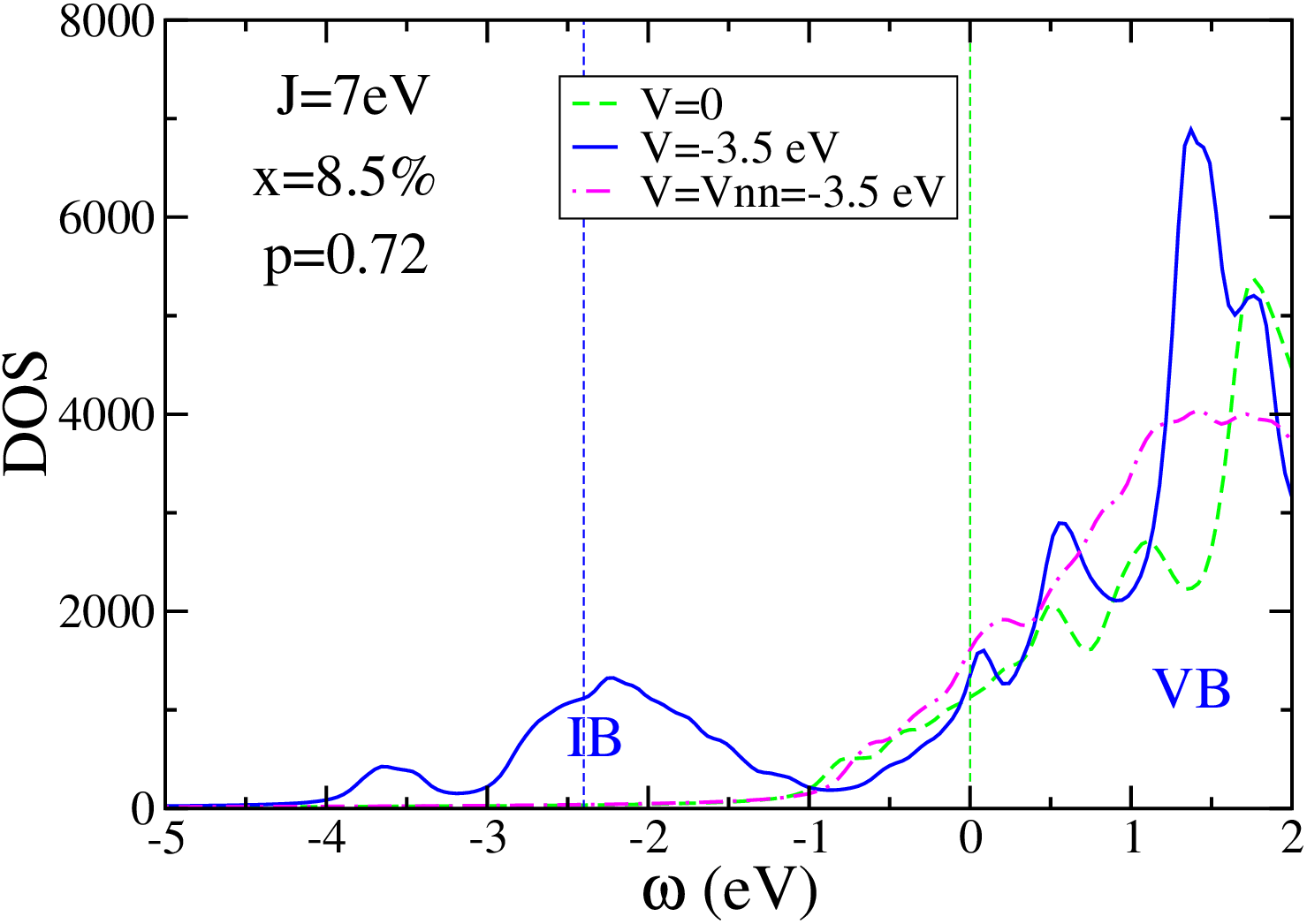}
\vskip 0.3cm
\caption{(color online) Numerically determined density of states for $J=7$eV 
with x=8.5\%.
For $V=0$ (dashed line), on site $V$=-3.5 eV (continuous line), and 
nearest-neighbor range $V$=-3.5 eV (dot-dashed line).}
\label{gaas}
\end{center}
\end{figure}

The above result indicates that studies of the crossover from impurity to 
valence band regimes performed with on-site rather than extended attractive 
potentials will overestimate the range of the impurity band regime.

\subsection{Mn $d$-orbitals}

Finally, in this subsection, we will consider the possibility of Mn$^{3+}$ 
doping in some III-V compounds such as GaN. 

\begin{figure}[thbp]
\begin{center}
\includegraphics[width=8cm,clip,angle=0]{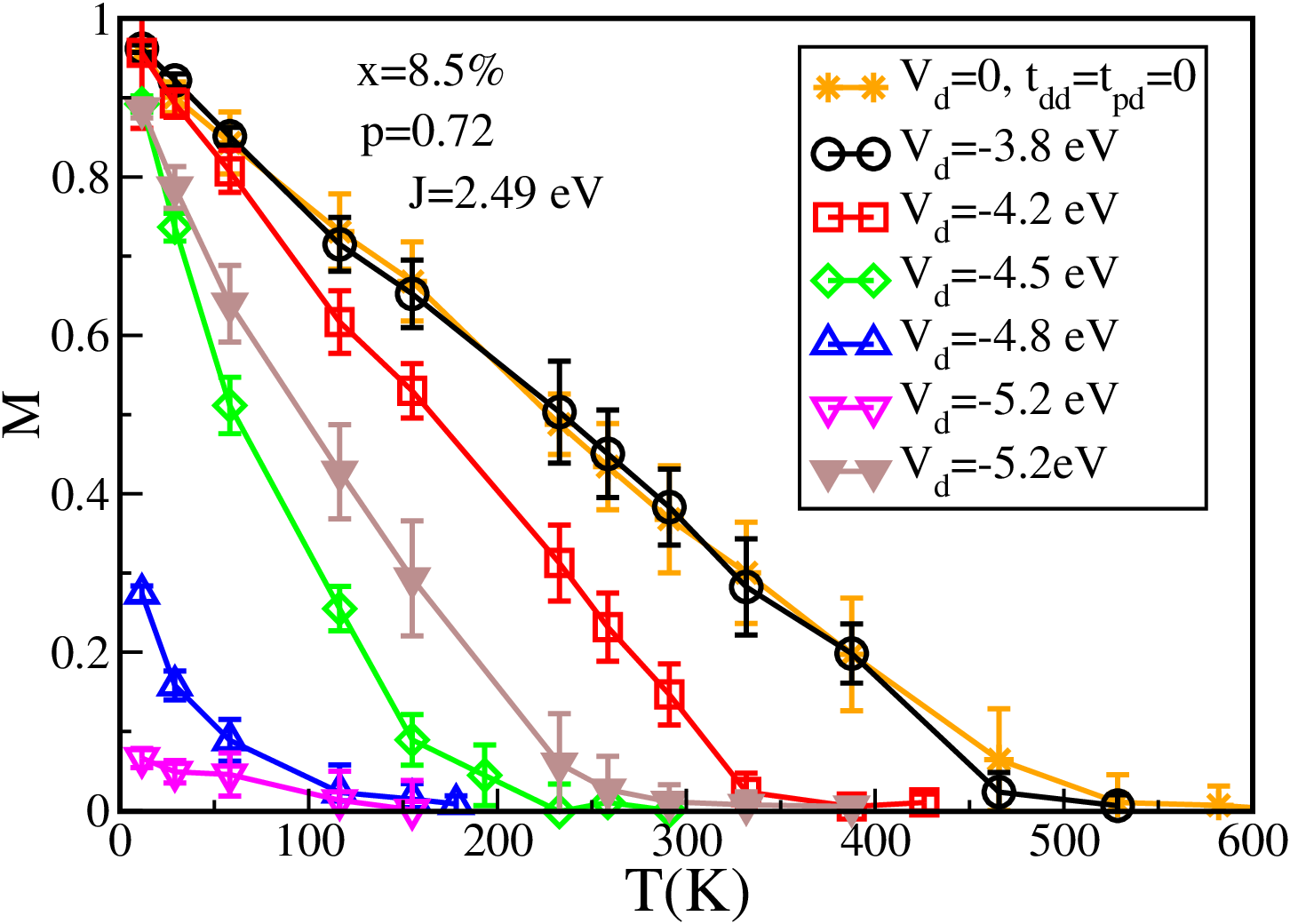}
\vskip 0.3cm
\caption{(color online) Numerically determined magnetization vs $T$ for Mn 
doped GaN with 
phenomenological $d$ orbitals. The open symbols correspond to $t_{dd}=-0.1eV$ 
and $t_{pd}=0.1eV$ while the close symbols indicate $t_{dd}=0.5eV$ and 
$t_{pd}=0.5eV$. The stars are data for the 6-orbital model.}
\label{gaas}
\end{center}
\end{figure}

In Fig.10 the stars indicate the magnetization as a function of temperature 
obtained using the six-orbital Hamiltonian in Eq.(31) with the parameters for 
GaN on a 256 sites lattice which has already been presented in Fig.5. For this 
material some experiments\cite{Graf,Kreis} and 
calculations\cite{Kro,Thomas,Kula} 
suggest that Mn $d$-levels may not be deep into the valence 
band. Thus, Mn$^{3+}$, instead of Mn$^{2+}$, may 
be present. To explore the consequences of this possibility we will
use the 8-orbital model that results from the 
addition of Eq.(34) to Eq.(31). As mentioned in Section V.B the relative 
position of the $d$ level $V_d$ and the $d-d$ and $d-p$ hoppings will be input 
parameters. First, we will consider the case in which the hoppings involving 
the $d$-orbitals are smaller than the hoppings among $p$-orbitals. Fixing 
$t_{dd}=-0.1$eV and $t_{pd}=0.1$eV, we present the magnetization vs T for 
various values of $V_d$ indicated in Fig.10 (open symbols). 
For $V_d=-3.8$eV, which 
corresponds to the $d$-orbitals deep into the valence band (open circles), the 
results are similar to the ones obtained for the 6-orbital case as expected.
However, as $V_d$ becomes more negative indicating that the $d$-level moves 
into 
the gap it can be clearly seen that both the magnetization and $T_C$ 
decrease. However, it is important to notice that, even in the case of 
$d$-orbitals in the gap, the magnetization and $T_C$ increase if the hoppings
between $d$ and $p$ orbitals are allowed to be larger. For example, for 
$|t_{pd}|=|t_{dd}|=0.5$eV and $V_d=-5.2$eV, indicated with filled triangles 
in Fig.10, the Curie temperature becomes very close to room temperature. This
behavior is in agreement with the {\it ab initio} results of 
Ref.\onlinecite{Kro}.

\begin{figure}[thbp]
\begin{center}
\includegraphics[width=8cm,clip,angle=0]{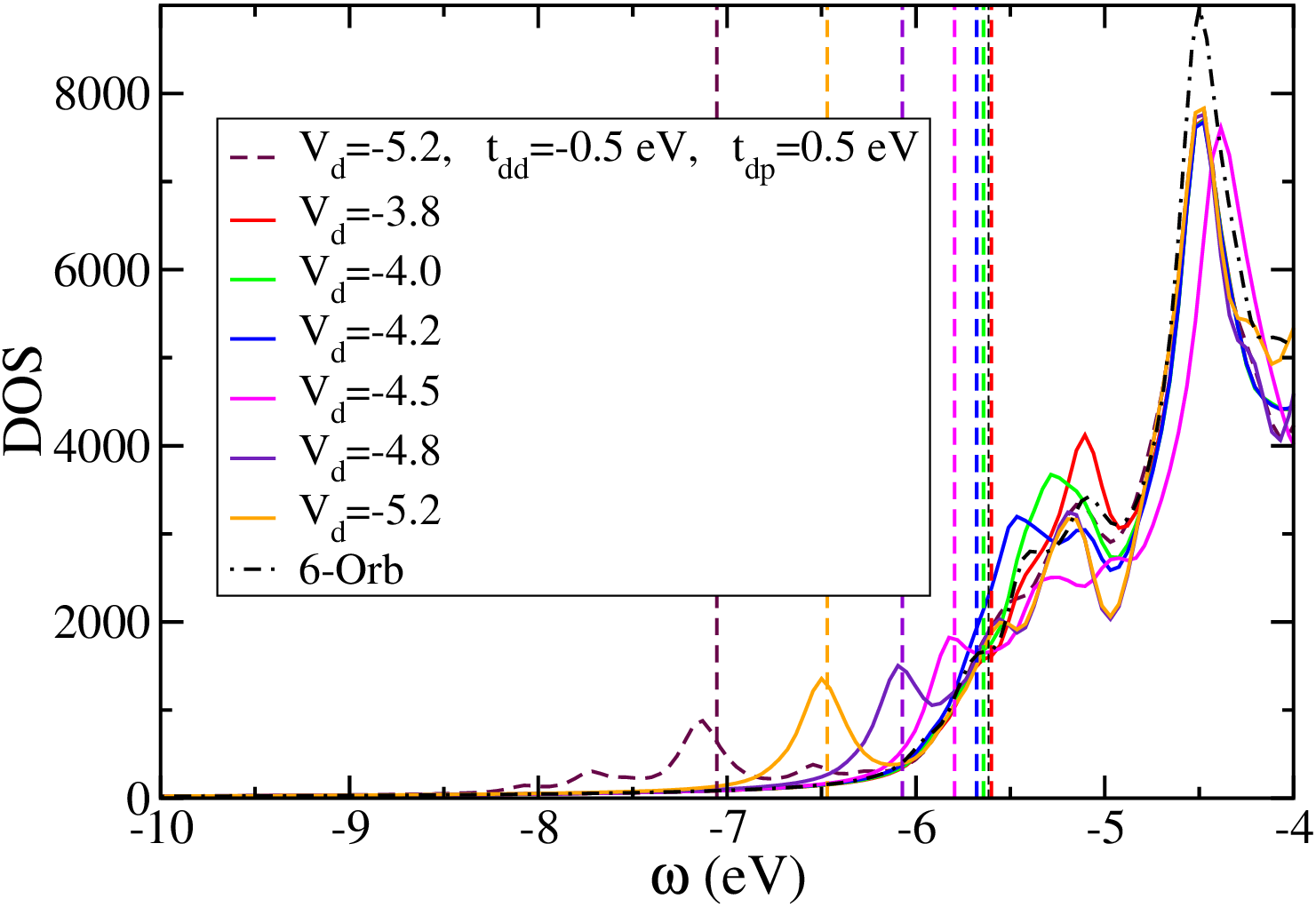}
\vskip 0.3cm
\caption{(color online) Numerically determined DOS for Mn doped GaN with 
phenomenological d orbitals. The continuous lines correspond to 
$t_{dd}=-0.1eV$ and $t_{pd}=0.1eV$ for the values of $V_d$ indicated while the 
dashed line indicates $t_{dd}=0.5eV$ and 
$t_{pd}=.5eV$. Data for the 6-orbital model are denoted with dashed-dotted 
line.}
\label{gaas}
\end{center}
\end{figure}

$T_C$ is related to the band structure as it can be seen in Fig.11. As $|V_d|$
becomes larger an impurity band develops in the DOS. When $t_{pd}<<t_{pp}$ the
IB has very small dispersion indicating that the trapped holes are very 
localized. When $t_{pd}$ increases the IB acquires dispersion and $T_C$ 
becomes higher. 

\section{Conclusions}

We have presented a Hamiltonian that enables the 
numerical study of the physics of DMS with an equally accurate 
treatment of both the kinetic and magnetic interactions. Provided that the 
parent 
compound is a Zinc-Blende type semiconductor for which the Luttinger 
parameters are known, there are no arbitrary parameters in the Hamiltonian. 
Only 6 orbitals per site have to be considered to obtain accurate Curie 
temperatures and magnetization curves of materials where Mn$^{2+}$ is 
present, which makes possible to perform numerical simulations in large 
systems with present day computers. Besides reproducing well 
established experimental results for GaAs with less than 10\% Mn doping, the 
higher doping regime was explored. While naively it could have been expected
that $T_C$ would increase continuously with $x$, we observed that it reaches a 
maximum value at an intermediate value of the doping. For GaAs we found that 
the highest value of $T_C$, approximately 220K, occurs $x$=12.5\% and no 
compensation, indicating that higher doping of GaAs will not lead to
$T_C$'s above room temperature. The doping for which the maximum $T_C$ occurs
appears to be in general agreement with recent experimental results.
 
The possibility of Mn$^{3+}$ doping in GaP or GaN has been considered 
phenomenologically. We found that $T_C$ will be very reduced if the $d$ levels 
are located at energies much higher than the top of the valence band and if 
the $pd$ hybridization is weak. However, if the hybridization is strong, Mn 
doped GaN could be a good candidate for high $T_C$.

It has also been shown that the Coulomb attraction between the hole and the 
Mn ions expected at low doping does not play a role for $x$ larger than 3\% in
GaAs. As a consequence, we have not observed an impurity band that should be 
present
at lower dopings or even in the metallic regime if the magnetic interaction 
were stronger.The relative weakness of the magnetic interaction was confirmed 
by studying the splitting of the top of the valence band as a function of $J$.
For values of $J$ within the experimental range for Mn doped GaAs, we have 
observed a departure from the mean-field predictions. However, the holes do 
not appear to be localized at the Mn impurity sites since the eigenvalue 
distribution is independent of the particular disorder configuration. 
Localization is observed only at unrealistically high values of $J$.   

Summarizing, we have shown that the real space Hamiltonian presented 
here reproduces the 
top of the valence band given by the Luttinger-Kohn model for the undoped 
parent compounds and, as such, it is 
guaranteed to provide an excellent framework for the numerical study of 
lightly doped Zinc-blende type magnetic semiconductors.

\section{Acknowledgments}

We acknowledge discussions with E. Dagotto, 
T. Schulthess, F. Popescu and J. Moreno. Y.Y. and A.M.
are supported by NSF under grants DMR-0706020.
This research used resources of the NCCS. G.A. is sponsored by the
Division of Scientific User Facilities and the Division of Materials Sciences 
and Engineering, BES, DOE, contract DE-AC05-000R22725 with ORNL, managed by 
UT-Battelle.

\section{Appendix I}

Here we provide the change of basis matrices used in Eq.(12) and (13).
$$
M=\left(\begin{array}{cccccc}
-{1\over{\sqrt{2}}}&-{i\over{\sqrt{2}}}&0&0&0&0 \\
{1\over{\sqrt{6}}}&-{i\over{\sqrt{6}}}&0&0&0&\sqrt{{2\over{3}}}\\
0 &0&\sqrt{{2\over{3}}}&{-1\over{\sqrt{6}}}&{-i\over{\sqrt{6}}} & 0 \\
0&0&0&{1\over{\sqrt{2}}}&-{i\over{\sqrt{2}}}&0 \\
0&0&-{1\over{\sqrt{3}}}&-{1\over{\sqrt{3}}}&-{i\over{\sqrt{3}}}&0 \\
-{1\over{\sqrt{3}}}&{i\over{\sqrt{3}}}&0&0&0&{1\over{\sqrt{3}}} 
\end{array} \right) 
\eqno(AI)
$$
\noindent and its inverse $M^{-1}$ is
$$
M^{-1}=\left(\begin{array}{cccccc}
-{1\over{\sqrt{2}}}&{1\over{\sqrt{6}}}&0&0&0&-{1\over{\sqrt{3}}} \\
{i\over{\sqrt{2}}}&{i\over{\sqrt{6}}}&0&0&0&{-i\over{\sqrt{3}}}\\
0 &0&\sqrt{{2\over{3}}}&0&{-1\over{\sqrt{3}}} & 0 \\
0&0&{-1\over{\sqrt{6}}}&{1\over{\sqrt{2}}}&{-1\over{\sqrt{3}}}&0 \\
0&0&{i\over{\sqrt{6}}}&{i\over{\sqrt{2}}}&{i\over{\sqrt{3}}}&0 \\
0&\sqrt{{2\over{3}}}&0&0&0&{1\over{\sqrt{3}}} 
\end{array} \right) 
\eqno(AII)
$$


\begin{thebibliography}{}
\bibitem{chang1} Similar expressions were obtained in Ref.\onlinecite{chang}.
\bibitem{foot} The sign of the diagonal hoppings determines whether the band is
hole-like or electron-like. The sign of the interorbital hopping $t_{xy}$ 
does not affect the shape of the band that only depends on the absolute 
value of this hopping.
\bibitem{chang2} These expressions are implicit in Eq.(6) of 
Ref.\onlinecite{chang}.
\bibitem{equiv} Notice that although $K$ and $U$ are equivalent points in 
terms of the FBZ the path between these two points and $\Gamma$ is not the 
same and the bands dispersion along the two directions are very different.
\end{thebibliography}
\end{document}